\documentclass[aps,twocolumn,nofootinbib,preprintnumbers,superscriptaddress,eqsecnum]{revtex4-1}

\usepackage[utf8]{inputenc}
\usepackage{color}
\usepackage[normalem]{ulem}
\usepackage{amsmath}
\usepackage{enumerate}
\usepackage{amsfonts}
\usepackage{epsfig}
\usepackage[
      colorlinks=true,
      linkcolor=blue,
      urlcolor=blue,
      filecolor=black,
      citecolor=red,
      pdfstartview=FitV,
      pdftitle={},
        pdfauthor={},
        pdfsubject={},
        pdfkeywords={},
        pdfpagemode=None,
        bookmarksopen=true
      ]{hyperref}
\usepackage{subcaption}      
      
\newcommand{\be}{\begin{equation}}
\newcommand{\ee}{\end{equation}}
\newcommand{\bea}{\begin{eqnarray}}
\newcommand{\eea}{\end{eqnarray}}

\newcommand\sP{{\ensuremath{{\mathcal P}}}}

\newcommand{\nn}{\nonumber\\}
\newcommand{\bln}{\begin{align}}
\newcommand{\eln}{\end{align}}
\newcommand{\bst}{\begin{split}}
\newcommand{\est}{\end{split}}
\newcommand{\bi}{\begin{itemize}}
\newcommand{\ei}{\end{itemize}}
\newcommand{\ben}{\begin{enumerate}}
\newcommand{\een}{\end{enumerate}}

\def\le{\left}
\def\ri{\right}
\def\ha{{1\over 2}}

\def\Lam{{\Lambda}}

\def\al{{\alpha}}

\def\th{{\theta}}

\def\Om{{\Omega}}
\def \th{{\theta}}

\def \om {\omega}

\def\sig{{\sigma}}
\def\ep{{\varepsilon}}

\newcommand{\p}{\partial}

\newcommand\ga{{\ensuremath{{\gamma}}}}
\newcommand\Ga{{\ensuremath{{\Gamma}}}}

\def\eeq{\end{equation}}

\newcommand\sL{{\ensuremath{{\mathcal L}}}}

\newcommand\sO{{\ensuremath{{\mathcal O}}}}

\newcommand\sW{{\mathcal W}}

\newcommand\CV{{\mathcal V}}

\newcommand\bJ{{\overline{J}}}

\begin{document}

\title {Generalized global symmetries and dissipative magnetohydrodynamics }

\author{Sa\v{s}o Grozdanov}
\email{grozdanov@lorentz.leidenuniv.nl}
\affiliation{Instituut-Lorentz for Theoretical Physics, Leiden University,\\Niels Bohrweg 2, Leiden 2333 CA, The Netherlands }
\author{Diego M. Hofman}
\email{d.m.hofman@uva.nl}
\author{Nabil Iqbal}
\email{n.iqbal@uva.nl}
\affiliation{Institute for Theoretical Physics, University of Amsterdam, Science Park 904, Postbus 94485, 1090 GL Amsterdam, The Netherlands}

\begin{abstract}
The conserved magnetic flux of $U(1)$ electrodynamics coupled to matter in four dimensions is associated with a generalized global symmetry. We study the realization of such a symmetry at finite temperature and develop the hydrodynamic theory describing fluctuations of a conserved 2-form current around thermal equilibrium. This can be thought of as a systematic derivation of relativistic magnetohydrodynamics, constrained only by symmetries and effective field theory. We construct the entropy current and show that at first order in derivatives, there are seven dissipative transport coefficients. We present a universal definition of resistivity in a theory of dynamical electromagnetism and derive a direct Kubo formula for the resistivity in terms of correlation functions of the electric field operator. We also study fluctuations and collective modes, deriving novel expressions for the dissipative widths of magnetosonic and Alfv\'en modes. Finally, we demonstrate that a non-trivial truncation of the theory can be performed at low temperatures compared to the magnetic field: this theory has an emergent Lorentz invariance along magnetic field lines, and hydrodynamic fluctuations are now parametrized by a fluid tensor rather than a fluid velocity. Throughout, no assumption is made of weak electromagnetic coupling. Thus, our theory may have phenomenological relevance for dense electromagnetic plasmas. 
\end{abstract}

\maketitle

\newpage
\begingroup
\hypersetup{linkcolor=black}
\tableofcontents
\endgroup

\section{Introduction}\label{intro}
Hydrodynamics is the effective theory describing the long-distance fluctuations of conserved charges around a state of thermal equilibrium. Despite its universal utility in everyday physics and its pedigreed history, its theoretical development continues to be an active area of research even today. In particular, the new laboratory provided by gauge/gravity duality has stimulated  developments in hydrodynamics alone, including an understanding of universal effects in anomalous hydrodynamics \cite{Son:2009tf, Neiman:2010zi}, potentially fundamental bounds on dissipation \cite{Kovtun:2004de,Son:2007vk}, a refined understanding of higher-order transport \cite{Baier:2007ix,Bhattacharyya:2008jc,Romatschke:2009kr,Bhattacharyya:2012nq,Jensen:2012jh,Grozdanov:2015kqa,Haehl:2014zda}, and path-integral (action principle) formulations of dissipative hydrodynamics \cite{Dubovsky:2011sj, Endlich:2012vt, Grozdanov:2013dba,Kovtun:2014hpa,Harder:2015nxa,Crossley:2015evo,Haehl:2015foa,Haehl:2015uoc,Torrieri:2016dko}; see e.g. \cite{Son:2007vk,Rangamani:2009xk,Kovtun:2012rj} for reviews of hydrodynamics from the point of view afforded by holography. 

It is well-understood that the structure of a hydrodynamic theory is completely determined by the conserved currents and the realization of such symmetries in the thermal equilibrium state of the system. In this paper we would like to apply such a symmetry-based approach to the study of magnetohydrodynamics, i.e. the long-distance limit of Maxwell electromagnetism coupled to light charged matter at finite temperature and magnetic field. 

To that end, we first ask a question with a seemingly obvious answer: what are the symmetries of $U(1)$ electrodynamics coupled to charged matter? One might be tempted to say that there is a $U(1)$ current $j_{\mathrm{el}}^{\mu}$ associated with electric charge. There is indeed such a divergenceless object, related to the electric field-strength by Maxwell's equations:
\be
\frac{1}{g^2} \nabla_{\mu} F^{\mu\nu} = j_{\mathrm{el}}^{\nu} . \label{emeom}
\ee
However, the symmetry associated with this current is a {\it gauge} symmetry. Gauge symmetries are merely redundancies of the description, and thus are presumably not useful for organizing universal physics. 

The true global symmetry of $U(1)$ electrodynamics is actually something different. Consider the following antisymmetric tensor
\be
J^{\mu\nu} = \ha \ep^{\mu\nu\rho\sig} F_{\rho\sig}.  \label{Jem}
\ee
It is immediately clear from the Bianchi identity (i.e. the absence of magnetic monopoles) that $\nabla_{\mu}J^{\mu\nu} = 0$. This is not related to the conservation of electric charge, but rather states that magnetic field lines cannot end.   

What is the symmetry principle behind such a conservation law? It has recently been stressed in \cite{Gaiotto:2014kfa} that just as a normal $1$-form current $J^{\mu}$ is associated with a global symmetry, higher-form symmetries such as $J^{\mu\nu}$ are associated with {\it generalized global symmetries}, and should be treated on precisely the same footing. We first review the physics of a conventional global symmetry, which we call a $0$-form symmetry in the notation of \cite{Gaiotto:2014kfa}: with every $0$-form symmetry comes a divergenceless $1$-form current $j^{\mu}$ , whose Hodge dual we integrate over a codimension-$1$ manifold to obtain a conserved charge. If this codimension-$1$ manifold is taken to be a time slice, then the conserved charge can be conveniently thought of as counting a conserved particle number: intuitively, since particle world-lines cannot end in time, we can ``catch'' all the particles by integrating over a time slice. The objects that are charged under $0$-form symmetries are local operators which create and destroy particles, and the symmetry acts (in the $U(1)$ case) by multiplication of the operator by a $0$-form phase $\Lam$ that is weighted by the charge of the operator $q$: $\sO(x) \to e^{iq \Lam} \sO(x)$.   

\begin{figure}[h]
\begin{center}
\includegraphics[scale=0.5]{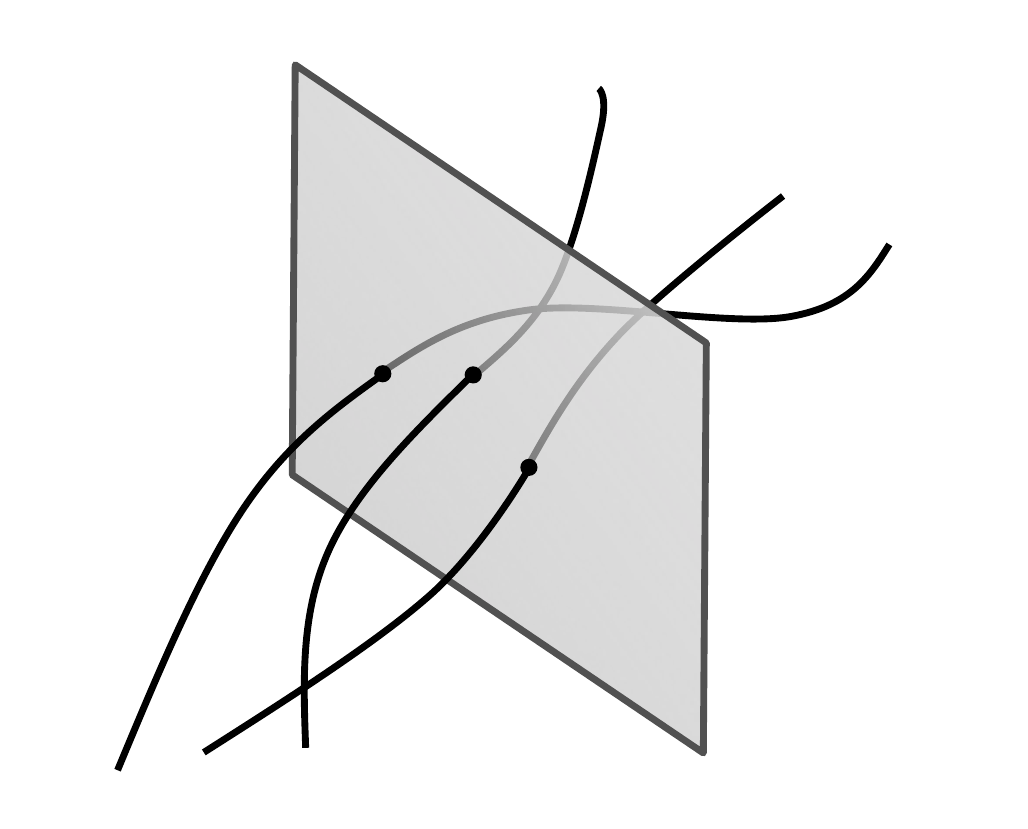}
\end{center}
\caption{Integration over a codimension-$2$ surface $S$ counts the number of strings that cross it at a given time.}
\label{fig:strings}
\end{figure}

Consider now the less familiar but directly analogous case of a $1$-form symmetry. A $1$-form symmetry comes with a divergenceless $2$-form current $J^{\mu\nu}$, whose Hodge star we integrate over a codimension-$2$ surface to obtain a conserved charge $Q = \int_{S} \star J$. This conserved charge should be thought of as counting a {\it string} number: as strings do not end in space {\it or} in time, an integral over a codimension-$2$ surface is enough to ``catch'' all the strings\footnote{Note that the dynamics of string-like degrees of freedom has been discussed in the context of superfluid hydrodynamics in the interesting recent paper \cite{Horn:2015zna}. In that case strings arise as solitons and, unlike in our work, interact through long range forces.}, as shown in Figure \ref{fig:strings}. The objects that are charged under $1$-form symmetries are $1$-dimensional objects such as Wilson or 't Hooft lines. These 1d objects create and destroy strings, and the symmetry acts (in the $1$-form case) by multiplication by a $1$-form phase $\Lam_{\mu}$ integrated along the contour $C$ of the 't Hooft line: $W(C) \to \exp\le(i q \int_C \Lam_{\mu} dx^{\mu}\ri) W(C)$. 

In the case of electromagnetism, the $2$-form current is given by \eqref{Jem}, and the strings that are being counted are magnetic field lines. We could also consider the dual current $F^{\mu\nu}$ itself, which would count {\it electric} flux lines: however from \eqref{emeom} we see that $F^{\mu\nu}$ is not conserved in the presence of light electrically charged matter, because electric field lines can now end on charges. Thus, electrodynamics coupled to charged matter has only a {\it single} conserved $2$-form current. This is the universal feature that distinguishes theories of electromagnetism from other theories, and the manner in which the symmetry is realized should be the starting point for further discussion of the phases of electrodynamics.\footnote{In electrodynamics in $2+1$ dimensions this point of view is somewhat more familiar, as the analog of $J^{\mu\nu}$ is a conventional $1$-form ``topological'' current $J^{\mu}_{2+1} = \ep^{\mu\nu\rho}F_{\nu\rho}$.} For example, this symmetry is spontaneously broken in the usual Coulomb phase (where the gapless photon is the associated Goldstone boson), and is unbroken in the superconducting phase (where magnetic flux tubes are gapped). We refer the reader to \cite{Gaiotto:2014kfa} for a detailed discussion of these issues. 

In this paper, we discuss the long-distance physics of this conserved current {\it near thermal equilibrium}, applying the conventional machinery of hydrodynamics to a theory with a conserved $2$-form current and conserved energy-momentum. We are thus constructing a generalization of the (very well-studied) theory that is usually called relativistic magnetohydrodynamics. To the best of our knowledge, most discussions of MHD separate the matter sector from the electrodynamic sector. It seems to us that this separation makes sense only at weak coupling, and may often not be justified: for example, the plasma coupling constant $\Ga$, defined as the ratio of potential to kinetic energies for a typical particle, is known to attain values up to $\mathcal{O}(10^2)$ in various astrophysical and laboratory plasmas \cite{RevModPhys.54.1017}. Experimental estimates of the ratio of shear viscosity to entropy density (where a small value is widely understood as being a universal measure of interaction strength \cite{Kovtun:2004de}) in such plasmas at high $\Ga$ obtain minimum values that are $\mathcal{O}(1)-\mathcal{O}(10)$ \cite{PhysRevLett.111.125004}. These suggest the presence of strong electromagnetic correlations. 


Our discussion will not make any assumptions of weak coupling and should therefore be valid for any value of $\Ga$: we will be guided purely by symmetries and the principles of the effective field theory of hydrodynamics. Beyond the (global) symmetries, the construction of the hydrodynamic gradient expansions also requires us to choose relevant hydrodynamic fields (degrees of freedom), which, as we will discuss, crucially depend on the symmetry breaking pattern in the physical system at hand. In particular, in addition to conventional hydrodynamics at finite temperature, we will also study a variant of magnetohydrodynamics at at very low temperatures. This theory has an emergent Lorentz invariance associated with boosts along the background magnetic field lines, and the parametrization of hydrodynamic fluctuations is considerably different. Interestingly, at $T = 0$ leading-order corrections to ideal hydrodynamics only enter at second order, thus showing the direct relevance of higher-order hydrodynamics (see e.g. \cite{Baier:2007ix,Bhattacharyya:2012nq,Romatschke:2009kr,Grozdanov:2015kqa}). While this treatment does not include the typical light modes that emerge at $T=0$, it does capture a universal self-contained sector of magnetohydrodynamics.   


We now describe an outline of the rest of this paper. In Section \ref{sec:ideal} we discuss the construction of  ideal hydrodynamic theory at finite temperature. In Section \ref{sec:1storder} we move beyond ideal hydrodynamics: we work to first order in derivatives, and demonstrate that there are seven transport coefficients that are consistent with entropy production, describing also how they may be computed through Kubo formulas. In Section \ref{sec:ModesFiniteT} we study linear fluctuations around the equilibrium solution and derive the dispersion relations and dissipative widths of gapless magnetohydrodynamic collective modes. In Section \ref{sec:baryon} we study the simple extension of the theory associated with adding an extra conserved $1$-form current (e.g. baryon number). In Section \ref{sec:T0} we turn to the theory at strictly zero temperature, where we discuss novel phenomena that can be understood as arising from a hydrodynamic equilibrium state with extra unbroken symmetries. We conclude with a brief discussion and possible future applications in Section \ref{sec:conc}. 

While this work was being written up, we came to learn of the interesting paper \cite{Schubring:2014iwa}, which also studies a dissipative theory of strings and makes the connection to MHD. Though the details of some derivations differ, there is overlap between that work and our Sections \ref{sec:ideal} and \ref{sec:1storder}. 

{\bf Note added:} In the original version of this work on the arXiv there was an inaccurate count of transport coefficients; we thank the authors of \cite{Hernandez:2017mch} for bringing this issue to our attention. 


\section{Ideal magnetohydrodynamics} \label{sec:ideal}

Our hydrodynamic theory will describe the dynamics of the slowly evolving conserved charges, which in our case are the stress-energy tensor $T^{\mu\nu}$ and the antisymmetric current $J^{\mu\nu}$. 
\subsection{Coupling external sources}
For what follows, it will be very useful to couple the system to external sources. The external source for the stress-energy tensor is a background metric $g_{\mu\nu}$, and we also couple the antisymmetric current $J^{\mu\nu}$ to an external $2$-form gauge field source $b_{\mu\nu}$ by deforming the microscopic on-shell action $S_0$ by a source term:
\begin{align}
S[b] \equiv S_0 + \Delta S[b], && \Delta S[b] \equiv \int d^4x\sqrt{-g}\;b_{\mu\nu} J^{\mu\nu} .\label{sourceb}
\end{align}
The currents are defined in terms of the total action as
\begin{align}
T^{\mu\nu}(x) &\equiv \frac{2}{\sqrt{-g}}\frac{\delta S}{\delta g_{\mu\nu}(x)} , \\ 
J^{\mu\nu}(x) &\equiv \frac{1}{\sqrt{-g}} \frac{\delta S}{\delta b_{\mu\nu}(x)}, \label{Jdefsource}
\end{align}
Demanding invariance of this action under the gauge symmetry $\delta_{\Lam} b = d\Lam$ with $\Lam$ a $1$-form gauge-parameter results in
\be
\nabla_{\mu} J^{\mu\nu} = 0 .  \label{consJ}
\ee
Similarly, demanding invariance under an infinitesimal diffeomorphism that acts on the sources as a Lie derivative $\delta_{\xi}g = \sL_{\xi}g$, $\delta_{\xi} b = \sL_{\xi} b$, gives us the (non)-conservation of the stress-energy tensor in the presence of a source:
\be
\nabla_{\mu} T^{\mu\nu} = H^{\nu}_{\phantom{\nu}\rho\sig}J^{\rho\sig}  , \label{consT}
\ee
where $H = db$. The term on the right-hand side of the equation states that an external source can perform work on the system. 

We now discuss the physical significance of the $b$-field source. A term $b_{ti} = \mu$ should be thought of as a chemical potential for the charge $J^{ti}$, i.e. a string oriented in the $i$-th spatial direction. 

For our purposes we can obtain some intuition by considering the theory of electrodynamics coupled to such an external source, i.e. consider using \eqref{Jem} to write the current as
\be
J^{\mu\nu} = \ep^{\mu\nu\rho\sig} \p_{\rho} A_{\sig} , \label{Afield}
\ee
with $A$ the familiar gauge potential from electrodynamics.\footnote{We choose conventions whereby $\ep_{txyz} = 1$.} Then the coupling \eqref{sourceb} becomes after an integration by parts:
\begin{align}
\Delta S[b] = \int d^4x \sqrt{-g} A_{\sig} j^{\sig}_{\mathrm{ext}} , && j^{\sig}_{\mathrm{ext}} \equiv \ep^{\sig\rho\mu\nu} \p_{\rho} b_{\mu\nu}, \label{jdef}
\end{align}
The field strength $H$ associated to $b$ can be interpreted as an {\it external} background electric charge density to which the system responds. 

For example, consider a cylindrical region of space $V$ that has a nonzero value for the chemical potential in the $z$ direction:
\be
b_{tz}(x) = \frac{\mu}{2}\theta_V(x), 
\ee
where $\theta_V(x)$ is $1$ if $x \in V$ and is $0$ otherwise. Then from \eqref{jdef} we see that we have 
\be
j^{\phi}_{\mathrm{ext}}(x) = \mu \delta_{\p V}(x) , 
\ee
i.e. we have an effective electric current running in a delta-function sheet in the $\phi$ direction along the outside of the cylinder. Thus the chemical potential for producing a magnetic field line poking through a system is an electrical current running around the edge of the system, as one would expect from textbook electrodynamics. In our formalism the actual magnetic field created by this chemical potential is controlled by a thermodynamic function, the susceptibility for the conserved charge density $J^{tz}$.

We will sometimes return to the interpretation of $b$ as charge source to build intuition: however we stress that in general when there are light electrically charged degrees of freedom present the $A(x)$ defined in \eqref{Afield} does not have a local effective action and is not a useful quantity to consider. 


\subsection{Hydrodynamic stress-energy tensor and current}
We now turn to ideal hydrodynamics at non-zero temperature. We first discuss the equilibrium state. Recall that the analog of a conserved charge $Q$ for our 2-form current is its integral over a codimension-$2$ spacelike surface $S$ with no boundaries, as shown in Figure \ref{fig:strings}.
\be
Q = \int_{S} \star J . 
\ee
$Q$ counts the number of field lines crossing $S$ at any instant of time and is thus unaltered by deformations of $S$ in both space and time. A thermal equilibrium density matrix is then given (for a particular choice of $S$) by
\be
\rho(T,\mu) = \exp\le(-\frac{1}{T}\le(H - \mu Q\ri)\ri),
\ee
where $\mu$ is the chemical potential associated with the $2$-form charge. This density matrix can be generated by a Euclidean path integral with an appropriate component of $b$ turned on, e.g. the $S$ is the $xy$ plane then we would use $b_{tz} = \frac{\mu}{2}$.  

Elementary arguments, which we spell out in detail in Appendix \ref{app:partition}, then give us the form of the stress-energy tensor and the conserved higher rank current in thermal equilibrium\footnote{Equilibrium thermodynamics in the presence of magnetic fields has also recently been studied in \cite{Kovtun:2016lfw}; that work differs from ours in that the magnetic fields there are fixed external sources for a conventional 1-form current, whereas in our case the magnetic fields are themselves the fluctuating degrees of freedom of a 2-form current.}:
\begin{equation}\label{idealTJ}
\begin{aligned}
T^{\mu\nu}_{(0)} & = (\ep + p)\, u^{\mu}u^{\nu} + p \, g^{\mu\nu} - \mu\rho\, h^{\mu}h^{\nu}, \\
J^{\mu\nu}_{(0)} & = 2\rho \, u^{[\mu}h^{\nu]} , 
\end{aligned}
\end{equation}
\noindent satisfying the conservation equations in the ideal limit

\bea
\nabla_\mu T^{\mu\nu}_{(0)} =0 \, , \quad\quad \nabla_\mu J^{\mu\nu}_{(0)}=0  \, .
\eea

We have labeled this expression with a subscript $0$, as this will be only the zeroth order term in an expansion in derivatives. Here $u^{\mu}$ is the fluid velocity as in conventional hydrodynamics. $h^{\mu}$ is the direction along the field lines, and we impose the following constraints:
\begin{align}\label{ConsT1}
u_\mu u^\mu = - 1, && h_\mu h^\mu = 1, && h_\mu  u^\mu = 0. 
\end{align}
It will also often be useful to use the projector onto the two dimensional subspace orthogonal to both $u$ and $h$:
\be\label{FiniteTProjector}
\Delta^{\mu\nu} = g^{\mu\nu} + u^{\mu}u^{\nu} - h^{\mu}h^{\nu},
\ee
with trace $\Delta^\mu_{~\mu} = 2$. In \eqref{idealTJ}, $\rho$ is the conserved flux density, and $p$ is the pressure. There is no mixed $u^{\mu} h^{\nu}$ term, as this can be removed with no loss of generality by a Lorentz boost in the $(u,h)$ plane.\footnote{We note that the form of the stress-energy tensor \eqref{idealTJ}, including constraints \eqref{ConsT1}, is precisely that of anisotropic ideal hydrodynamics with different longitudinal and transverse pressures (with respect to some vector) \cite{Ryblewski:2008fx,Florkowski:2008ag}. In that case, $\mu\rho$ measures the difference between the two pressures. The role of this additional vector is now played by $h^\mu$.} 

Note the presence of the $h^{\mu}h^{\nu}$ term in the stress-energy tensor, representing the tension in the field lines. Its coefficient in equilibrium is $\mu\rho$. It is a bit curious from the effective field theory perspective that this coefficient is fixed and is not given by an equation of state, like $p$, for example. There is a quick thermodynamic argument to explain this fact. Consider the variation of the internal energy for a system containing field lines running perpendicularly to a cross section of area $A$, with an associated tension $\tau$ and a conserved charge $Q$ given by the flux through the section:
\be
dU = T dS - p dV + \tau A dL + \mu L dQ ,
\ee
\noindent where $L$ is the length of the system perpendicular to $A$. Because $Q$ is a charge defined by an area integral, it is given by $Q=\rho A$ and the factor of $L$ in front of $dQ$ is the correct scaling with the height of the system. Now perform a Legendre transform to the Landau grand potential:
\begin{align}
\Phi &= U - T S - \mu L Q , \\
 d\Phi &= - s V dT - p dV - \rho V d \mu + \left(\tau -\mu \rho \right) A dL  , \quad
\end{align}
\noindent where $s$ is the entropy density. Notice that $\Phi$ is the quantity naturally calculated by the on-shell action and we expect it to scale with volume in local Quantum Field Theory. This scaling is spoiled by the term proportional to $dL$ unless $\tau =\mu \rho$. This condition is, therefore, enforced by extensivity.

The thermodynamics is, thus, completely specified by a single equation of state, i.e. by the pressure as a function of temperature and chemical potential $p(T, \mu)$. The relevant thermodynamic relations are
\begin{align}
\ep + p = Ts + \mu\rho, && dp = s dT + \rho d\mu , \label{thermo}
\end{align}
with $s$ the entropy density.  Here we have made use of the volume scaling assumption.

The microscopic symmetry properties of $J$ do not actually determine those of $h^{\mu}$ and $\rho$, only that of their product. In this work we assume the charge assignments in Table \ref{tbl:disc}, which are consistent with magnetohydrodynamical intuition and are particularly convenient. Note that that all scalar quantities (such as $\rho$ and $\mu$) are taken to have even parity under all discrete symmetries, and charge conjugation is taken to flip the sign of $h$. These symmetries will play a useful role later on in restricting corrections to the entropy current. 
\begin{table}
\begin{tabular}{c|c | c | c| c | c | c|c}

& $J_{ti}$ & $J_{ij}$ & $u_t$ & $u_i$ & $h_t$ & $h_i$ & $\rho,\mu,\ep,p$ \\
\hline 
$C$ & $-$ &$-$ &$+$ &$+$& $-$& $-$& $+$ \\  \hline
$P$ & $+$ & $-$ & $+$ & $-$ & $-$ & $+$ & $+$ \\ \hline
$T$ & $-$ & $+$ & $+$ & $-$ & $+$ & $-$ & $+$  \\ \hline 
\end{tabular}
\caption{Charges under discrete symmetries of 2-form current and hydrodynamical degrees of freedom.} 
\label{tbl:disc}
\end{table}

Hydrodynamics is a theory that describes systems that are in local thermal equilibrium but can globally be far from equilibrium, in which case the thermodynamic degrees of freedom become space-time dependent hydrodynamic fields. Thus the degrees of freedom are the two vectors $u^{\mu}, h^{\mu}$ and two thermodynamic scalars which can be taken to be $\mu$ and $T$, leading to seven degrees of freedom. The equations of motion are the conservation equations \eqref{consT} and \eqref{consJ}. As $J$ is antisymmetric, one of the equations for the conservation of $J$ does not include a time derivative and is a constraint on initial data. This constraint is consistently propagated by the remaining equations of motion, thus leaving effectively six equations for six variables, and the system is closed. 


We now demonstrate that the equations of motion of ideal hydrodynamics result in a conserved entropy current. Consider dotting the velocity $u$ into the conservation equation for the stress-energy tensor \eqref{consT}. Using the thermodynamic identities \eqref{thermo} we find
\begin{align}
u_{\nu} \nabla_{\mu} T^{\mu\nu} =& -T \nabla_{\mu}\le(s u^{\mu}\ri) - \mu \p_{\mu}\le(\rho u^{\mu}\ri)  \nn
& - \mu\rho\le(u_{\nu} \nabla_{\mu} h^{\nu}\ri)h^{\mu} = 0 . \label{div1}
\end{align}
We now project the conservation equation for $J$ along $h^{\mu}$:
\be
h_{\nu}\nabla_{\mu} J^{\mu\nu} = \nabla_{\mu}\le(\rho u^{\mu}\ri) - \rho h^{\mu}\le(\nabla_{\mu}u^{\nu}h_{\nu}\ri) = 0  . \label{consJh}
\ee
Inserting this into \eqref{div1} and using $\nabla_{\mu}(u^{\nu} h_{\nu}) = 0$ to rearrange derivatives we find
\be
\nabla_{\mu} (s u^{\mu}) = 0  . 
\ee
We thus see that the local entropy current $s u^{\mu}$ is conserved, as we expect in ideal hydrodynamics. 

We now turn to the interpretation of the other components of the hydrodynamic equations. The projections of \eqref{consT} along $h_{\nu}$ and $\Delta_{\nu\sig}$, respectively, are
\begin{align}
h_\nu\left[  (\ep + p)u^{\mu}\nabla_{\mu} u^{\nu} + \nabla^{\nu}p \right] - \nabla_{\mu}(\mu\rho h^{\mu}) & = 0 ,  \\
\Delta_{\nu\sig}\le[ (\ep + p)u^{\mu}\nabla_{\mu} u^{\nu} + \nabla^{\nu}p - \mu\rho h^{\mu}\nabla_{\mu}h^{\nu}\ri] & = 0 .
\end{align}
These are the components of the Euler equation for fluid motion in the direction parallel and perpendicular to the background field. 

Similarly, the evolution of the magnetic field is given by the projection of the conservation equation for $J^{\mu\nu}$ along $h_{\nu}$ in \eqref{consJh} and along $\Delta_{\nu\sig}$ below:
\be
\Delta_{\nu\sig}\le(u^{\mu} \nabla_{\mu} h^{\nu} - h^{\mu} \nabla_{\mu} u^{\nu}\ri) = 0  . 
\ee
The equation states that the transverse part of the magnetic field is Lie dragged by the fluid velocity. 

This is the most general system that has the symmetries of Maxwell electrodynamics coupled to charged matter. In particular, unlike conventional treatments of MHD, we have made no assumption that the $U(1)$ gauge coupling $g^2$ is weak. Indeed it appears nowhere in our equations: in theories with light charged matter, the fact that $g^2$ runs means that it does not have a universal significance and will not appear as a fundamental object in hydrodynamic equations. 

To make contact with the traditional treatments of MHD, consider expanding the pressure in powers of $\mu$, e.g.
\be\label{WeakEqStateQED}
p\left(\mu, T\right) = p_0(T) + \frac{1}{2}g(T)^2 \mu^2 + \cdots.
\ee 
Here $p_0(T)$ should be thought of as the pressure of the matter sector alone. The expansion is given in powers of $\mu^2$, as the sign of $\mu$ is not physical\footnote{In this theory, the sign of the magnetic field is carried by the direction of the $h^\mu$ vector.}. If we stop at this order and then further assume that the coefficient of the $\mu^2$ term is independent of temperature $g(T) = g$, then the theory of ideal hydrodynamics arising from this particular equation of state is entirely equivalent to traditional relativistic MHD with gauge coupling given by $g$. From our point of view, this is then a weak-magnetic-field limit of our more general theory. Note that this weak-field limit is entirely different from the hydrodynamic limit that we are taking throughout this paper, and there is an entirely consistent effective theory even if we do not take the weak-field limit. We discuss some physical consequences of keeping higher order terms in this expansion (which will be generically present in any interacting theory, even if their coefficient may be small under particular circumstances) later on in this paper.  

Nevertheless, if we truncate the expansion for the pressure as in \eqref{WeakEqStateQED} then we find from \eqref{thermo}: $\rho = g^2\mu$ and $\ep = \ep_0 + \frac{\rho^2}{2 g^2}$ with $\ep_0 = T\partial_T p_0(T) - p_0$. The ideal hydrodynamic theory of our $2$-form current is now entirely equivalent to conventional treatments of ideal MHD, as presented in e.g. \cite{PhysRevD.18.1809}. As $s \sim \p_T p$, the $T$-independence of $g$ and thus of the $\mu$-dependent piece of the pressure essentially means that the magnetic field degrees of freedom carry no entropy.

\section{First-order hydrodynamics} \label{sec:1storder}

Hydrodynamics is an effective theory, and thus \eqref{idealTJ} are only the zeroth order terms in a derivative expansion. We now move on to first order in derivatives: to be more precise, the full stress-energy tensor is given by
\begin{align}
T^{\mu\nu} & = T^{\mu\nu}_{(0)} + T^{\mu\nu}_{(1)} + \cdots , \\
J^{\mu\nu} & = J^{\mu\nu}_{(0)} + J^{\mu\nu}_{(1)} + \cdots, 
\end{align}
where the zeroth order term is given by the ideal MHD expressions in \eqref{idealTJ}, and our task now is to determine the first-order corrections as a function of the fluid variables such as the velocity and magnetic field. The numbers that parametrize these corrections are the transport coefficients such as viscosity and resistivity. The physics of dissipation and entropy increase enter at first order in the derivative expansion: as usual in hydrodynamics, the possible tensor structures that can appear (and thus the number of independent transport coefficients) are greatly constrained by the requirement that entropy always increases.

\subsection{Transport coefficients}
We follow the standard procedure to determine these corrections \cite{landau1987fluid}. We begin by writing down the most general form for the first-order terms:
\begin{align}
T^{\mu\nu}_{(1)} & = \delta \ep\, u^{\mu}u^{\nu} + \delta f \,\Delta^{\mu\nu} + \delta \tau\, h^{\mu}h^{\nu} + 2 \,\ell^{(\mu}h^{\nu)} \nonumber \\ & + 2 \,k^{(\mu}u^{\nu)} + t^{\mu\nu} ,  \nonumber \\ 
J^{\mu\nu}_{(1)} & = 2\, \delta\rho \,u^{[\mu}h^{\nu]} + 2 \,m^{[\mu}h^{\nu]} + 2 \, n^{[\mu}u^{\nu]} + s^{\mu\nu}. \label{1stordercorr}
\end{align}
Here $\ell^{\mu}, k^{\mu}, m^{\mu}$, and $n^{\mu}$ are transverse vectors (i.e. orthogonal to both $u^\mu$ and $h^\mu$), $t^{\mu\nu}$ is a transverse, traceless and symmetric tensor, and $s^{\mu\nu}$ is a transverse, antisymmetric tensor. 

Next, we exploit the possibility to change the hydrodynamical frame. In hydrodynamics, there is no intrinsic microscopic definition of the fluid variables $\{ u^{\mu}, h^{\mu},\mu, T\}$. Each field can therefore be infinitesimally redefined, as e.g. $u^{\mu}(x) \to u^{\mu}(x) + \delta u^{\mu}(x)$. The microscopic currents and the stress-energy tensor must remain invariant under this operation, and thus the redefinition alters the functional form of the relationship between the currents and the fluid variables. In conventional hydrodynamics of a charged fluid this freedom is often used to set $T^{\mu\nu}_{(1)}u_{\nu} = 0$ (Landau frame) or $j^{\mu}_{(1)} = 0$ (Eckart frame). We will use the scalar redefinitions of $\mu$ and $T$ to set $\delta \rho = \delta \ep = 0$ and the vector redefinitions of $u^\mu$ and $h^\mu$ to set $k^{\mu} = n^{\mu} = 0$. We now have the simpler expansion:

\begin{align}
T^{\mu\nu}_{(1)} & = \delta f \, \Delta^{\mu\nu} + \delta \tau \, h^{\mu}h^{\nu}  + 2 \, \ell^{(\mu}h^{\nu)} + t^{\mu\nu} , \label{corrT} \\ 
J^{\mu\nu}_{(1)} & =  2 \, m^{[\mu}h^{\nu]} + s^{\mu\nu} . \label{corr}
\end{align}
Our task now is to determine the form of the reduced set $\{ \delta f, \delta \tau, \ell^{\mu}, m^{\mu},t^{\mu\nu}, s^{\mu\nu}\}$ in terms of derivatives of the fluid variables. 

To proceed, we require an expression for the non-equilibrium entropy current $S^{\mu}$. The textbook approach to this problem is to postulate a standard ``canonical'' form for this entropy current, motivated by promoting the thermodynamic relation $Ts = p + \ep - \mu \rho$ to the following covariant expression:
\be\label{CovThermoT}
T S^{\mu} = p \, u^{\mu} - T^{\mu\nu} u_{\nu} - \mu  \, J^{\mu\nu} h_{\nu} . 
\ee
Up to first order in derivatives, this is equivalent to
\be
S^{\mu} = s \,u^{\mu} - \frac{1}{T}T^{\mu\nu}_{(1)}u_{\nu}  - \frac{\mu}{T} J^{\mu\nu}_{(1)}h_{\nu} \label{canS}
\ee
We will take this to be our entropy current. As in conventional hydrodynamics \cite{Bhattacharya:2011eea}, one can show that it is invariant under frame redefinitions of the sort described above. 

Next, we directly evaluate the divergence $\nabla_{\mu} S^{\mu}$. Using the contraction of the conservation equations \eqref{consT} and \eqref{consJ} with $u^{\mu}$, we find after some straightforward algebra:
\begin{align}\label{EntProd1stOrd}
\nabla_{\mu} S^{\mu}  =&\, -\biggr[ T^{\mu\nu}_{(1)}\nabla_{\mu}\le(\frac{u_{\nu}}{T}\ri)  \nn
& \,+ J^{\mu\nu}_{(1)}\le(\nabla_{\mu}\le(\frac{h_{\nu}\mu}{T}\ri) + \frac{u_{\sig} H^{\sig}_{\phantom{\sig}\mu\nu}}{T}\ri)\biggr].
\end{align}
We see that entropy is no longer conserved, as one expects for a dissipative theory. The second law of thermodynamics in its local form states that entropy should always increase.  Thus the right-hand side of Eq. \eqref{EntProd1stOrd} should be a positive definite quadratic form for all conceivable fluid flows. 
For the vector and tensor dissipative terms, positivity implies that the right-hand side is simply a sum of squares, requiring that the dissipative corrections take the following form:
\begin{align}
 \ell^{\mu} & = -2\eta_{\parallel}\Delta^{\mu\sig}h^{\nu} \nabla_{(\sig}u_{\nu)} ,\label{visc4} \\
t^{\mu\nu} & = -2\eta_{\perp}\le(\Delta^{\mu\rho}\Delta^{\nu\sig}- \ha \Delta^{\mu\nu}\Delta^{\rho\sig}\ri)\nabla_{(\rho}u_{\sig)} ,\\
m^{\mu} & = -2 r_{\perp}\Delta^{\mu\beta}h^{\nu}\le( T \nabla_{[\beta}\le(\frac{h_{\nu]} \mu}{T}\ri) + u_{\sig} H^{\sig}_{\phantom{\sig}\beta\nu}\ri) ,\label{mdef}\\
s^{\mu\nu} & = -2 r_{\parallel}\Delta^{\mu\rho} \Delta^{\nu\sigma} \le(\mu \nabla_{[\rho} h_{\sigma]} +H^{\lambda}_{\phantom{\sig}\rho\sigma}u_{\lambda}\ri). \label{sdef},
\end{align}
where the four transport coefficients $\eta_{\perp,\parallel}$ and $r_{\perp,\parallel}$ must all be positive. 

In the bulk channel parametrized by $\delta f$ and $\delta \tau$ mixing is possible. The most general allowed form that is consistent with positivity is parametrized by three transport coefficients $\zeta_{\perp,\parallel,\times}$: 
\begin{align}
\delta f & = -\zeta_{\perp} \Delta^{\mu\nu} \nabla_{\mu}u_{\nu} - \zeta_{\times} h^{\mu}h^{\nu} \nabla_{\mu}u_{\nu} \label{bulk1} \\
\delta \tau & = -\zeta_{\times} \Delta^{\mu\nu} \nabla_{\mu} u_{\nu} - \zeta_{\parallel} h^{\mu} h^{\nu} \nabla_{\mu} u_{\nu} \label{viscfin}
\end{align}
Note that this mixing matrix is symmetric, in that the mixing term $\zeta_{\times}$ is the same for $\delta f$ and $\delta \tau$. This follows from an Onsager relation on mixed correlation functions, as we explain in Section \ref{sec:Kubo} below.\footnote{In the first version of this paper on the arXiv, the possibility of a nonzero $\zeta_{\times}$ was not taken into account, leading to an incorrect count of transport coefficients. This inaccuracy was pointed out to us by the authors of \cite{Hernandez:2017mch}, and we thank them for bringing this to our attention.}

Further demanding that the right-hand side of \eqref{EntProd1stOrd} be a positive-definite quadratic form imposes two constraints on the bulk viscosities, which may be written as
\be
\zeta_{\perp} \geq 0 \qquad \zeta_{\perp}\zeta_{\parallel} \geq \zeta_{\times}^2
\ee
There are no further constraints that we know of. At first order we thus have seven transport coefficients $\zeta_{\perp,\parallel,\times}$, $\eta_{\perp,\parallel}$ and $r_{\perp,\parallel}$. If we were to allow all coefficients permitted by symmetries, we would instead have concluded that there were eleven independent transport coefficients consistent with the parity assignments under $h^\mu \rightarrow - h^\mu$, illustrating the constraints enforced by the second law of thermodynamics.  


We now turn to the interpretation of these transport coefficients. It is clear that $\zeta_{\perp,\parallel,\times}$ and $\eta_{\perp,\parallel}$ are anisotropic bulk and shear viscosities respectively: for a charged fluid in a fixed external magnetic field one finds instead seven independent viscosities \cite{Huang:2011dc}, where the difference in counting arises from the fact that we have imposed a charge conjugation symmetry $h^{\mu} \rightarrow -h^{\mu}$.

The transport coefficients $r_{\parallel,\perp}$ can be interpreted as the conventional electrical resistivity parallel and perpendicular to the magnetic field. To understand this, first note that the familiar electric field $E^{\mu}$ is defined in terms of the electromagnetic field strength as $E^{\mu} = F^{\mu\nu}u_{\nu}$. Using \eqref{Jem} we find
\begin{align}
E_{\mu} &= -\ha \ep_{\mu\nu\rho\sig} u^{\nu}J^{\rho\sig} \nn
&= -\ha \ep_{\mu\nu\rho\sig} u^{\nu}\le(2 m^{[\rho}h^{\sig]} + s^{\rho\sig} + \cdots\ri), \label{Efield}
\end{align}
where the ellipsis indicates further higher-order corrections. Note that a nonzero electric field enters only at first order in hydro: an electric field is not a low-energy object, as the medium is attempting to screen it.

Next, we note that a resistivity is conventionally defined as the electric 1-form current response to an applied external electric field. However, our formalism instead naturally studies the converse object, i.e. the 2-form current response $J^{\mu\nu}$ in a field theory with a total action $S[b]$ deformed by a fixed external $b$-field source (which can be interpreted as an external electric current via \eqref{jdef}). Thus, we need to perform a Legendre transform to find the analog of the quantum effective action $\Ga[\bJ]$, which is a function of a specified 2-form current $\bJ$: 
\be
\Ga[\bJ] \equiv S[b] - \int d^{4}x \sqrt{-g}\;b_{\mu\nu} \bJ^{\mu\nu} . 
\ee
Here, $S[b]$ is defined to be the on-shell action in the presence of the $b$-field source, and $b$ is implicitly determined by the condition that $J \equiv \delta_b S = \bJ$, i.e. that the stationary points of the action coincide with the specified value for $\bJ$. We now write $\bJ$ in terms of a vector potential $\overline{A}$ using \eqref{Afield} and define the electrical 1-form current {\it response} $\overline{j}^{\mu}$ via 
\be
\overline{j}^{\mu}(x) \equiv \frac{\delta \Ga[\bJ]}{\delta \overline{A}_{\mu}(x)} = -\ep^{\mu\nu\rho\sig} \p_{\nu} b_{\rho\sig} . \label{response}
\ee  
Note the sign difference with respect to the external fixed source $j^{\mu}_{\mathrm{ext}}$ defined in \eqref{jdef}. This arises from the Legendre transform and is the difference between having a fixed external source and a current response. 

We now need to determine the relationship between the electric field \eqref{Efield} and the response 1-form current \eqref{response}. Consider a static and homogenous fluid flow with 
\begin{align}
u^{\mu}(x) = \delta^{\mu}_t , && h^{\mu}(x) = \delta^\mu_z, \label{staticflow}
\end{align}
in the presence of a homogenous but time-dependent $b$ field source $b_{xy}(t)$, $b_{xz}(t)$. 
From \eqref{response}, in the fixed $\overline{J}$ ensemble, this $b$-field can be interpreted as an electrical current response $\overline{j}^{z} = -2 \dot{b}_{xy},\;\overline{j}^{y} = 2 \dot{b}_{xz}.$ Now inserting the expansion  \eqref{mdef} and \eqref{sdef} into \eqref{Efield} and neglecting the fluid gradient terms, we find that the electric field created by this current source is
\begin{align}
E_z = r_{\parallel} \overline{j}^z , && E_y = r_{\perp} \overline{j}^y .
\end{align}
Thus, $r_{\parallel,\perp}$ are indeed anisotropic resistivities as claimed. 

Finally, we discuss a technical point: our starting point for the discussion of dissipation was the canonical form for the non-equilibrium entropy current \eqref{canS}. It is now well-understood that this form for the entropy current is not unique: for example, in the hydrodynamics of fluids with anomalous global symmetries (and thus with parity violation), the second law {\it requires} that extra terms must be added to the entropy current, resulting eventually in extra transport coefficients corresponding to the chiral magnetic and vortical effects \cite{Son:2009tf,Neiman:2010zi}. It was however shown in \cite{Bhattacharya:2011tra} that for a parity-preserving fluid with a conserved $1$-form current all ambiguities in the entropy current can be fixed by demanding that entropy production on an arbitrary curved background be positive. We have performed a similar analysis for the $2$-form current. Here, charge-conjugation invariance acts as $h^{\mu} \to -h^{\mu}$, and this symmetry together with positivity of entropy production on curved backgrounds is sufficient to show that the form of the entropy current exhibited in \eqref{canS} is unique. 

\subsection{Kubo formulae}\label{sec:Kubo} 
We now derive Kubo formulae---i.e. expressions in terms of real-time correlation functions---for these transport coefficients. We follow an approach described in \cite{Son:2007vk} which we briefly review below. 

It is a standard result in linear response theory that in the presence of a perturbation out of equilibrium by an infinitesimal source, the response is given of the system is given by the retarded correlator of the operator coupled to the source. For example, if we turn on a small $b$-field source, we find: 
\be
\delta \langle J^{\mu\nu}(\om,k) \rangle = -G_{JJ}^{\mu\nu,\rho\sig}(\om,k) b_{\rho\sig}(\om,k),
\ee
where $G_{JJ}^{\mu\nu,\rho\sig}(\om,k)$ is the retarded correlator of $J$. 

However, above we saw that in the presence of an infinitesimal perturbation around a static flow \eqref{staticflow} by a time-varying but spatially homogenous $b$-field source $b_{xy}(t)$, $b_{xz}(t)$, the response within the hydrodynamic theory was
\begin{align}
J^{xy} = -2 r_{\parallel} \dot{b}_{xy}(t) , && J^{xz} = -2 r_{\perp} \dot{b}_{xz}(t) .
\end{align}
Equating these two relations we find the following Kubo formulas for the parallel and perpendicular resistivities:
\begin{align}
r_{\parallel} = \lim_{\om \to 0} \frac{G_{JJ}^{xy,xy}(\om) }{-i\om} , && r_{\perp} = \lim_{\om \to 0} \frac{G_{JJ}^{xz,xz}(\om)}{-i\om}. \label{resKubo}
\end{align}

We will return to the physical interpretation of this formula shortly. First, we derive Kubo formulas for the viscosities. To do this, we consider perturbing the spatial part of the background metric slightly away from flat space:
\be
g_{ij} = \delta_{ij} + h_{ij}(t), \qquad g_{ti} = 0, \qquad g_{tt} = -1. 
\ee
where $h_{ij} \ll 1$. The response of the stress-energy tensor to such a perturbation is given in linear response theory by
\be
\delta\left \langle T^{ij}(\om,k) \right\rangle = -\ha G^{ij,kl}_{TT}(\om,k) h_{kl}(\om,k)  .
\ee
The hydrodynamic response to such a source is given by \eqref{visc4} to \eqref{viscfin} where the full contribution comes from the Christoffel symbol
\be
\nabla_{(i}u_{j)} = -\Gamma^t_{ij} u_t = \ha \dot{h}_{ij} . 
\ee
Matching the response in each tensor channel just as above we find the following set of Kubo relations:
\begin{align}
& \eta_{\parallel} = \lim_{\om \to 0} \frac{G_{TT}^{xz,xz}(\om) }{-i\om} , & \eta_{\perp} = \lim_{\om \to 0} \frac{G_{TT}^{xy,xy}(\om)}{-i\om} ,   \label{KuboShVis} \\
& \zeta_{\parallel} = \lim_{\om \to 0} \frac{G_{TT}^{zz,zz}(\om) }{- i\om} , & \zeta_{\perp} + \eta_{\perp} = \lim_{\om \to 0} \frac{G_{TT}^{xx,xx}(\om)}{-i\om}, 
\end{align}
\be
\zeta_{\times} = \lim_{\om \to 0} \frac{G_{TT}^{zz,xx}(\om)}{-i\om} = \lim_{\om \to 0}\frac{G_{TT}^{xx,zz}(\om)}{-i\om}. \label{KuboBulkVisX}
\ee
These are a straightforward anisotropic generalization of the usual formulas for the bulk and shear viscosity. Our normalization for the anisotropic bulk viscosity has been chosen so that no dimension-dependent factors enter into the Kubo formula; however this is not the standard normalization. Note that we present two equivalent formulas for the mixed bulk viscosity $\zeta_{\times}$; the equality of these two correlation functions is guaranteed by the Onsager relations for off-diagonal correlation functions. Indeed, it is this Onsager relation that sets to zero a possible antisymmetric transport coefficient in \eqref{bulk1}--\eqref{viscfin}.\footnote{The Kubo formulae \eqref{resKubo} and \eqref{KuboShVis}--\eqref{KuboBulkVisX} agree with those presented in \cite{Hernandez:2017mch}. We thank Pavel Kovtun for discussions regarding these matters.}

We now turn to a discussion of the resistivity formula \eqref{resKubo}. Unlike the hydrodynamics of a conventional $1$-form current where we generally obtain a Kubo formula for the {\it conductivity}, here we find a Kubo formula directly for its inverse, the resistivity, in terms of correlators of the components of the antisymmetric tensor current corresponding to the electric field. The resistivity is the natural object here: in a theory of dynamical electromagnetism, we examine how an electric field responds to an external current flow, not the other way around. 

To the best of our knowledge, such a Kubo formula for the resistivity in terms of electric field correlations is novel. Traditionally, in order to compute a resistivity one instead computes the conductivity of the $1$-form global current that is being gauged, and then takes the inverse of the resulting number ``by hand''. This procedure---which essentially treats a gauge symmetry as a global one---is probably only physically reasonable at weak gauge coupling. On the other hand, the Kubo formula above permits a precise universal definition for the resistivity in a dynamical $U(1)$ gauge theory, independently of the strength of the gauge coupling. It is interesting to study its implications. 

For example, we might see whether it agrees with the traditional prescription. Consider a weakly coupled $U(1)$ gauge theory with action
\be
S[A,\phi] = \int d^4 x\le(\frac{1}{g^2} F^2 + A_{\mu} j^{\mu}_{el}[\phi]\ri),
\ee
where $j^{\mu}_{el}$ is a $1$-form current that is built out of other matter fields (schematically denoted by $\phi$), that has been weakly gauged. The considerations here do not involve the background magnetic field and so we turn it to zero. Within this theory we may compute the finite-temperature correlator of the electric field to compute the resistivity through \eqref{resKubo}. 

\begin{figure}[h]
\begin{center}
\includegraphics[scale=0.5]{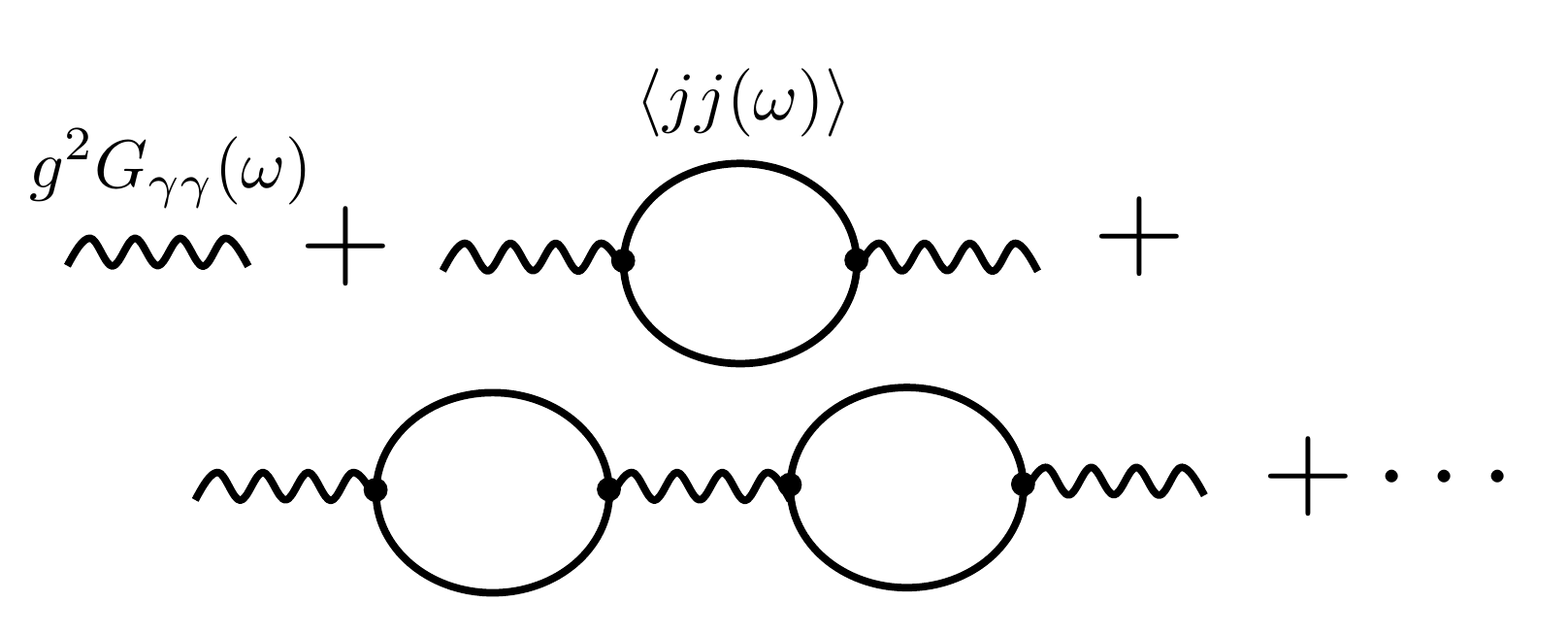}
\end{center}
\caption{Sum over current-current insertions to compute electrical resistivity.}
\label{fig:bubblesum}
\end{figure}

One first attempt to do so might involve summing the series of diagrams shown in Figure \ref{fig:bubblesum}. The geometric sum leads to an answer of the schematic form
\be
\left\langle E E(\om)\right \rangle \sim -(-i\om)^2 \frac{g^2 G_{\ga\ga}(\om)}{1 - \left\langle j_{el}j_{el} (\om) \right\rangle g^2 G_{\ga\ga}(\om)},
\ee
where $G_{\ga\ga}$ is the free photon propagator for spatial polarizations and $\langle j_{el} j_{el} \rangle$ is the correlation function of the electrical current. The photon propagator at zero spatial momentum has a pole at $\om \to 0$: at low frequencies we now zoom in on this pole to find for the resistivity $r$:
\be
r \sim (-i\om) \frac{1}{\langle j_{el}j_{el} (\om) \rangle} \sim \frac{1}{\sig},
\ee
where we have used the standard Kubo formula for the $0$-form global conductivity $\langle j_{el}j_{el} (\om \to 0) \rangle= -i\om \sigma$. Thus, {\it within this approximation scheme}, it is indeed true that the resistivity (defined via our Kubo formula) is equal to the inverse of the  conductivity of the current that is being gauged.\footnote{Here we have been somewhat cavalier with details. To make these considerations precise, one should imagine performing the sum over bubbles in Euclidean signature, then analytically continuing to the retarded propagator at frequency $\om$ via $\om_E \to -i\om$ before taking the small frequency limit. We have assumed here that no subtleties arise in this continuation.}

\begin{figure}[h]
\begin{center}
\includegraphics[scale=0.5]{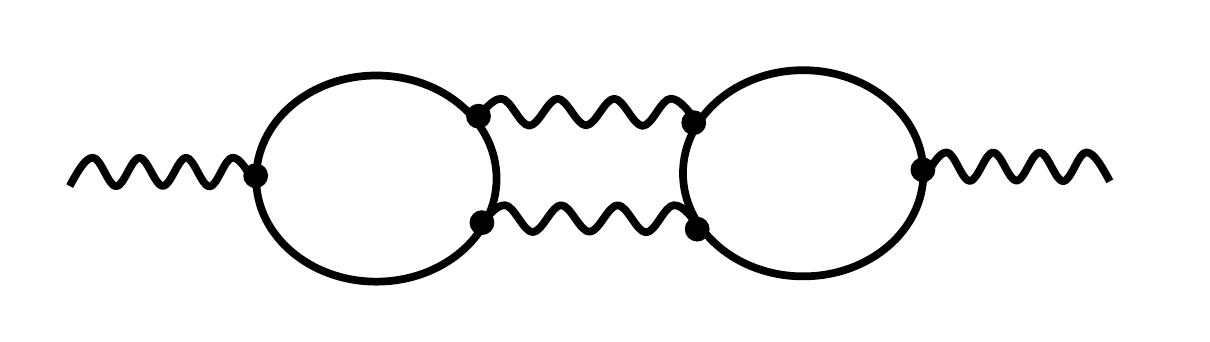}
\end{center}
\caption{Example of new diagram that contributes to electrical resistivity.}
\label{fig:twobubb}
\end{figure}
 
Note, however, that this class of diagrams is {\it not} the only set of diagrams that one should include. One might also imagine diagrams of the form Figure \ref{fig:twobubb}: computationally they arise from the fact that the photon is now dynamical, and thus the classification of diagrams as ``one-particle-irreducible'' has changed. Such diagrams will contribute to \eqref{resKubo}: as they simply do not exist in the theory of the global $1$-form current $j_{el}$, they will necessarily modify the conclusion above, changing $r$ away from $\sig^{-1}$. We have not attempted a systematic study of such diagrams, but it would be very interesting to understand their effect. It seems likely that they can be suppressed at weak gauge coupling, justifying the approximation scheme above, but it is an important open issue to demonstrate precisely when this is possible.    

\section{Application: Dissipative Alfv\'en and magnetosonic waves}\label{sec:ModesFiniteT}

In this section, we study the collective modes of the relativistic MHD theory constructed above. We will linearly perturb the background solution and determine the dispersion relations $\om(k)$ of the resulting modes. We organize the fluctuations in the following way: without loss of generality, we fix the direction of the background magnetic field by setting the $h^\mu$ field to point in the $z$-direction, $h^\mu = (0,0,0,1)$ (note that its size is fixed by the normalization of $h^\mu$). Furthermore, we can use a residual $SO(2)$ symmetry to fix the 4-momentum as
\begin{equation}
k^\mu = (\omega ,q ,0,k) \equiv (\omega, \kappa \sin\theta,0,\kappa\cos\theta),
\end{equation}
 so that $\theta$ measures the angle between the direction of the background magnetic field and momentum of the hydrodynamic waves. The background velocity field is fixed to $u^\mu = (1,0,0,0)$ at rest and the background temperature and chemical potential are kept general and space-time independent. We then linearly perturb $u^\mu$, $h^\mu$, $T$ and $\mu$ as
\begin{eqnarray}
& u^\mu & \to u^\mu + \delta u^\mu \,e^{- i\omega t + i q x + i k z}  ,\\
& h^\mu &\to h^\mu + \delta h^\mu \,e^{- i\omega t + i q x + i k z} ,\\
&T & \to T + \delta T \, e^{- i\omega t + i q x + i k z} ,\\
&\mu &\to \mu + \delta \mu \, e^{- i\omega t + i q x + i k z}  .
\end{eqnarray}
Note that linearized constraints \eqref{ConsT1} impose that
\begin{equation}
u_\mu \delta u^\mu =0  \, , \quad h_\mu \delta h^\mu =0 \, , \quad u_\mu \delta h^\mu + h_\mu \delta u^\mu=0\,.
\end{equation}
For a background source without curvature, i.e. $H^\mu_{\phantom{\mu}\rho\sigma} = 0$, the fluctuations can be organized into two classes:

\begin{itemize}
\item Transverse Alfv\'{e}n waves with
\begin{eqnarray}
&h_\mu \delta u^\mu & = u_\mu \delta h^\mu = 0 ,\\
&k_\mu \delta u^\mu &= k_\mu \delta h^\mu = 0 , \\
&\delta T & = \delta \mu =0.
\end{eqnarray}
Note that the fluid displacement is perpendicular to the background magnetic field; thus, they can be thought of as the usual vibrational modes that travel down a string with tension. These modes were first discovered in the magnetohydrodynamic context by Alfv\'{e}n in \cite{1942Natur.150..405A}. For an introductory treatment, see e.g. \cite{bellan2008fundamentals}.

 \item Magnetosonic waves with $\delta u^\mu$ and $\delta h^\mu$ contained in the space spanned by $\{u^\mu, h^\mu, k^\mu \}$. These are more closely related to the usual sound mode in a finite temperature plasma. We will see that there are two branches of this kind: ``fast" and ``slow".
\end{itemize} 
We first study Alfv\'en waves. Solving the conservation equations \eqref{consJ} and \eqref{consT}, we find the dispersion relation for Alfv\'{e}n waves to $\mathcal{O}(\kappa^2)$ to be
\begin{align}\label{DRAlfven}
\omega = \pm v_A  \kappa & - \frac{i}{2}  \bigg( \frac{1}{\ep + p}\le(\eta_\perp \sin^2 \th  +  \eta_\parallel \cos^2\theta\right) \nonumber \\ 
& + \frac{\mu}{\rho}\le( r_\perp \cos^2\theta + r_\parallel \sin^2\theta\right)\bigg) \kappa^2 ,
\end{align}
where the parameter that enters the Alfv\'{e}n phase velocity is
\begin{align}\label{AlfvenSpeed}
v_A^2 = \CV_A^2 \cos^2 \theta, && \CV_A^2=  \frac{\mu \rho}{\ep + p} .
\end{align}
The expression for the speed of the wave is standard. Recall that $\mu\rho$ is the tension in the field lines; in the nonrelativistic limit $(\ep + p)$ is dominated by the rest mass, and this becomes the textbook formula for the speed of wave propagation down a string. We are not however aware of much previous discussion of dissipative corrections to Alf\'{v}en waves; \cite{Jedamzik:1996wp} studied a dissipative fluid perturbatively coupled to electrodynamics, and our expression reduces to their angle-independent result if we assume an isotropic shear viscosity and no resistivity. 

When the magnetic field is perpendicular to the direction of momentum, i.e. $\cos^2 \theta = 0$, the Alfv\'{e}n wave ceases to propagate and becomes entirely diffusive, as is usually the case for transverse excitations in standard hydrodynamics. Note that the width of the mode depends on the momentum perpendicular to the strings; elementary treatments of MHD often assume that the Alfv\'{e}n wave has no dependence on the perpendicular momentum at all, which is sometimes taken as license to make it arbitrarily high, allowing Alfv\'{e}n waves that are arbitrarily well-localized in the plane perpendicular to the field (see e.g. \cite{bellan2008fundamentals}). Here, we see that this is an artefact of the ideal hydrodynamic limit. 

Turning now to the magnetosonic waves, a straightforward but somewhat tedious calculation shows that the dispersion relations for the two magnetosonic waves are given by
\begin{align}\label{DRMagS}
\omega = \pm v_M \kappa - i \tau  \,\kappa^2 ,
\end{align}
where 
\begin{align}
&v_{M}^2 = \frac{1}{2} \biggr\{ \left(\CV_A^2 + \CV_0^2 \right) \cos^2 \theta+ \CV_S^2 \sin^2 \theta \label{MSSpeed}\\
&  \pm  \sqrt{ \left[\left(\CV_A^2- \CV_0^2\right) \cos^2 \theta+ \CV_S^2 \sin^2 \theta  \right]^2  +  4 \CV^4 \cos^2\theta \sin^2\theta }   \biggr\} .\nonumber
\end{align}
Note that ``fast" magnetosonic waves have a $+$ sign before the square-root in Eq. \eqref{MSSpeed} and ``slow" magnetosonic waves have a $-$ sign. Above, we have defined the following quantities:
\begin{eqnarray}
\CV_0^2 &=&  \frac{ s \chi}{T (c \chi-\lambda^2)} , \\
\CV_S^2 &=& \frac{s^2 \chi + \rho^2 c-2 \rho s \lambda }{(c \chi-\lambda^2) (\ep + p)}, \\
\CV^4 &=& \frac{s (\rho \lambda - s \chi)^2}{T (c \chi - \lambda^2)^2 (\ep + p)} ,
\end{eqnarray}
and the susceptibilities:
\begin{equation}
\chi = \frac{\partial \rho}{\partial \mu} \, ,\quad c = \frac{\partial s}{\partial T} \, , \quad \lambda =  \frac{\partial s}{\partial \mu} =  \frac{\partial \rho}{\partial T}  \,. 
\end{equation}

It is easy to see that the formulae above predict generically the existence of a two fully dissipative modes at $\theta=\frac{\pi}{2}$, namely the ``slow" magnetosonic mode and the Alfv\'{e}n mode. We can interpret $\CV_S$ as the speed of the ``fast" magetosonic mode at $\theta = \frac{\pi}{2}$, a kind of speed of sound for the system. At $\theta=0$, on the other hand, one magnetosonic mode has the same speed as the Alfv\'{e}n mode while the other one has velocity $\CV_0$. We plot these velocities as a function of the angle $\theta$ for some interesting examples below.

The dissipative parts of these modes can be calculated in a straightforward manner by going to one higher order in derivatives using the formalism above. Unfortunately, explicit expressions are rather cumbersome to write in print. We quote below only the values for $\tau$ at $\theta=0$ and $\theta=\frac{\pi}{2}$, where we indicate which mode the width applies to by specifying the value of the phase velocity at that angle\footnote{Note that depending on the equation of state and the specific values of $\mu$ and $T$ (which determine the relative numerical magnitudes of $\CV_A$ and $\CV_0$) it can be either the fast or the slow magnetosonic mode that has phase velocity coinciding with the Alfv\'{e}n wave at $\th = 0$, as can be seen explicitly in Figures \ref{fig:highT} and \ref{fig:lowT}.}:
\begin{align}
\tau \left(\CV_A, \theta=0\right) &= \frac{1}{2} \left(\frac{\eta_\parallel}{\ep + p} + \frac{r_\perp \, \mu}{\rho} \right) , \label{disa2} \\
\tau \left(\CV_0, \theta=0\right) &= \ha \frac{\zeta_\parallel}{s T} , \\
\tau \left(0, \theta=\frac{\pi}{2}\right) &= \frac{1}{2} \left( \frac{\eta_\parallel}{s T} + \frac{r_\perp (\ep + p)^2}{T^2 (s^2 \chi + \rho^2 c - 2 \rho s \lambda)} \right), \\
\tau \left(\CV_S, \theta=\frac{\pi}{2}\right) &= \frac{1}{2} \left( \frac{\zeta_\perp + \eta_\perp}{\ep +p}\right. \nonumber\\
&+ \left. \frac{r_\perp ( c T \rho + \rho \lambda \mu - s T \lambda - s \mu \chi)^2}{T^2(c \chi - \lambda^2) (s^2 \chi + \rho^2 c - 2 \rho s \lambda)} \right).
\end{align}

While the coefficient $\zeta_{\times}$ enters into the dispersion relations of magnetosonic waves, its coefficient is proportional to $\sin^2\theta \cos^2 \theta$, which implies that the magnetosonic dispersion relations have neither any dependence on the bulk viscosity $\zeta_{\times}$ at $\theta = 0$ nor at $\theta = \pi / 2$. Notice that the dissipative part (\ref{disa2}) coincides exactly with the $\theta \rightarrow 0$ limit of (\ref{DRAlfven}). This is expected, as in this limit there is an enhanced $SO(2)$ rotational symmetry around the shared axis of background magnetic field and momentum, relating the modes in question. As a result of this coincidence the results presented allow the measurement of only 5 of the 7 dissipative coefficients. As it turns out, if we allow measurements at arbitrary angles, then $\zeta_{\times}$ can be determined, but the value of $\eta_\perp$ can't be measured from the study of dissipation of linear modes alone. By introducing sources, one can of course use the Kubo formulae previously discussed to determine all transport coefficients.

\subsection{Magnetohydrodynamics at weak field}

In order to recover the familiar results from standard magnetohydrodynamics, we can take the small chemical potential limit, which corresponds to weak magnetic fields. This is the regime in which the standard treatment is valid. 

In the weak field limit, we can expand the equation of state as (cf. \eqref{WeakEqStateQED})
\begin{align}
p_{\mathrm{weak}}\left(\mu, T\right) =&\,\, p_0(T) + \frac{1}{2}  g^2(T) \mu^2 +\cdots , 
\end{align}
\noindent where $p_0(T)$ and $g(T)$ are temperature-dependent functions that control the leading order behavior. In this limit, to leading order, 
\bea
v_A^2 &=& \frac{g^2 \mu^2}{s T} \cos^2 \theta + \cdots, \\
\left(v_M^2\right)_{\mathrm{fast}} &=& \frac{s}{c T} + \cdots, \\
\left(v_M^2\right)_{\mathrm{slow}} &=& \frac{g^2 \mu^2}{s T} \cos^2 \theta+ \cdots.
\eea

This agrees with the standard treatment (for a relativistic discussion, see e.g. \cite{Jedamzik:1996wp}). Notice that the slow magnetosonic mode and the Alfv\'{e}n wave are indistinguishable to this order. If we want to separate them we need to go to higher order in the expansion. One nice example when one can do this and obtain concrete expressions is in the case where $\mu$ is much larger than any other scale in the problem (while still being much smaller than $T^2$). In this case, we have no other scale and the expansion of the equation of state to the necessary order is:
\begin{align}\label{pweak}
p_{\mathrm{weak}}\left(\mu, T\right) =&\,\, \frac{a}{4} T^4 + \frac{g^2}{2} \mu^2 + \frac{\beta}{4}  \frac{\mu^4}{T^4} +\cdots ,
\end{align}
\noindent where $a$, $g$ and $\beta$ are dimensionless constants. We find the leading $\mu^2$ effects on the velocities of modes to be
\bea
v_A^2&=& \frac{g^2 \mu^2}{a T^4} \cos^2 \theta + \cdots, \\
\left(v_M^2\right)_{\mathrm{fast}} &=& \frac{1}{3} + \frac{2}{3} \frac{g^2 \mu^2}{a T^4} \sin^2 \theta + \cdots, \\
\left(v_M^2\right)_{\mathrm{slow}} &=& \frac{g^2 \mu^2}{a T^4} \cos^2 \theta+ \cdots, \\
 v_A^2 -\left(v_M^2\right)_{\mathrm{slow}} &=&  \frac{g^4 + a \beta}{2 a^2}\frac{ \mu^4}{T^8} \sin^2 2\theta+\cdots .\quad \quad
\eea
In each of the expressions, we have kept only the first non-trivial term to illustrate the angular dependence. The factor of $\frac{1}{3}$ in the leading order expression for $\left(v_M^2\right)_{\mathrm{fast}}$ is characteristic of the sound mode of conformal fluids in 4 dimensions. The fact that sound is the fastest mode is in agreement with our expectations at high temperatures where propagation is by nature diffusive. Note that both the Alfv\'en and the slow magnetosonic wave speeds start at $\sO(\mu^2)$, which is the small expansion parameter in this limit. Thus, they propagate very slowly indeed. We present some illustrative plots of these dispersion relations in Figure \ref{fig:highT}.

\subsection{Magnetohydrodynamics at strong field} \label{Tlessmu}

The situation is quite different for a fluid in which magnetic fields are strong. Here, our formalism can make concrete predictions away from the weak coupling limit. For concreteness let us assume, similarly as in the previous discussion, that $T^2$ is much larger than any other scale in the problem, while still much smaller than $\mu$. In that case we can write the equation of state in a small temperature expansion (strong magnetic field) as:
\be
p_{\mathrm{strong}}\left(\mu, T\right) = \frac{g'^2}{2} \mu^2 + \frac{a'}{4}  T^4 + \frac{\beta'}{8}  \frac{T^8}{\mu^2} +\cdots , \label{pstrong}
\ee
\noindent where $g'$, $a'$ and $\beta'$ are dimensionless constants. The expansion above is shown to the second subleading order to highlight that this expansion is, despite similarities, indeed different from (\ref{pweak}). The fact that the leading order terms agree (in form, but not numerical coefficients) between the two expansions is a coincidence due to our working in $4$ dimensions.

From the above equation of state, we can calculate the mode velocities to first non-trivial order in temperature corrections:
\begin{align}
v_A^2 & =  \, \cos^2 \theta-\frac{a' T^4}{g'^2 \mu^2} \cos^2 \theta + \cdots , \label{muexp1} \\
\left( v_M^2 \right)_{\mathrm{fast}} & = \, 1 -  \frac{a' T^4}{g'^2 \mu^2} \frac{2}{2+ \sin^2 \theta} + \cdots , \\
\left( v_M^2 \right)_{\mathrm{slow}}& = \, \frac{1}{3} \cos^2 \theta  \label{muexp2} \\
&- \frac{T^4 \cos^2\theta}{9 g'^2 a' \mu^2} \left(4 g'^2 \beta' +\frac{3 a'^2  \sin^2 \theta}{2 + \sin^2\theta}\right)   + \cdots. \nonumber
\end{align}

There are a few interesting features of these expressions. For propagation in the direction of the magnetic field lines, the Alfv\'{e}n wave now has the same velocity as the fast magnetosonic mode, instead of having the same velocity as the slow mode, which was the case in the large temperature expansion. Furthermore, the speed of these modes is that of light in the strict $T \rightarrow 0$ limit. This feature is completely general and independent of the particular no-scale assumption for the Alfv\'{e}n wave.  Another important difference is that Alfv\'{e}n modes can only propagate along magnetic field lines while fast magnetic modes propagate in any direction.

The slow magnetosonic mode is somewhat peculiar. Notice that the $\beta'$ coefficient contributes at an earlier order in $T$ than in the other modes. A more striking related feature is that the leading factor of $\frac{1}{3}$ is not universal and depends strongly on the power of the leading temperature contribution. If, for example, $a'$ had been zero, we would have found that the zero temperature velocity squared of the slow mode in the direction of the magnetic field lines was instead $\frac{1}{7}$. Therefore, the high magnetic field limit is non-universal for this mode. The reason behind this is that this mode is the only one that contains $\delta T$ fluctuations as $\mu \gg T^2$. It is an interesting question weather a universal hydrodynamic theory can be built that only describes the physics of fluctuating chemical potentials at fixed temperature if $\mu \gg T^2$. We answer this question in the affirmative in section \ref{sec:T0}.

\begin{figure*}[ht]
\begin{subfigure}{0.4\linewidth}
\centering
\includegraphics[width=\linewidth]{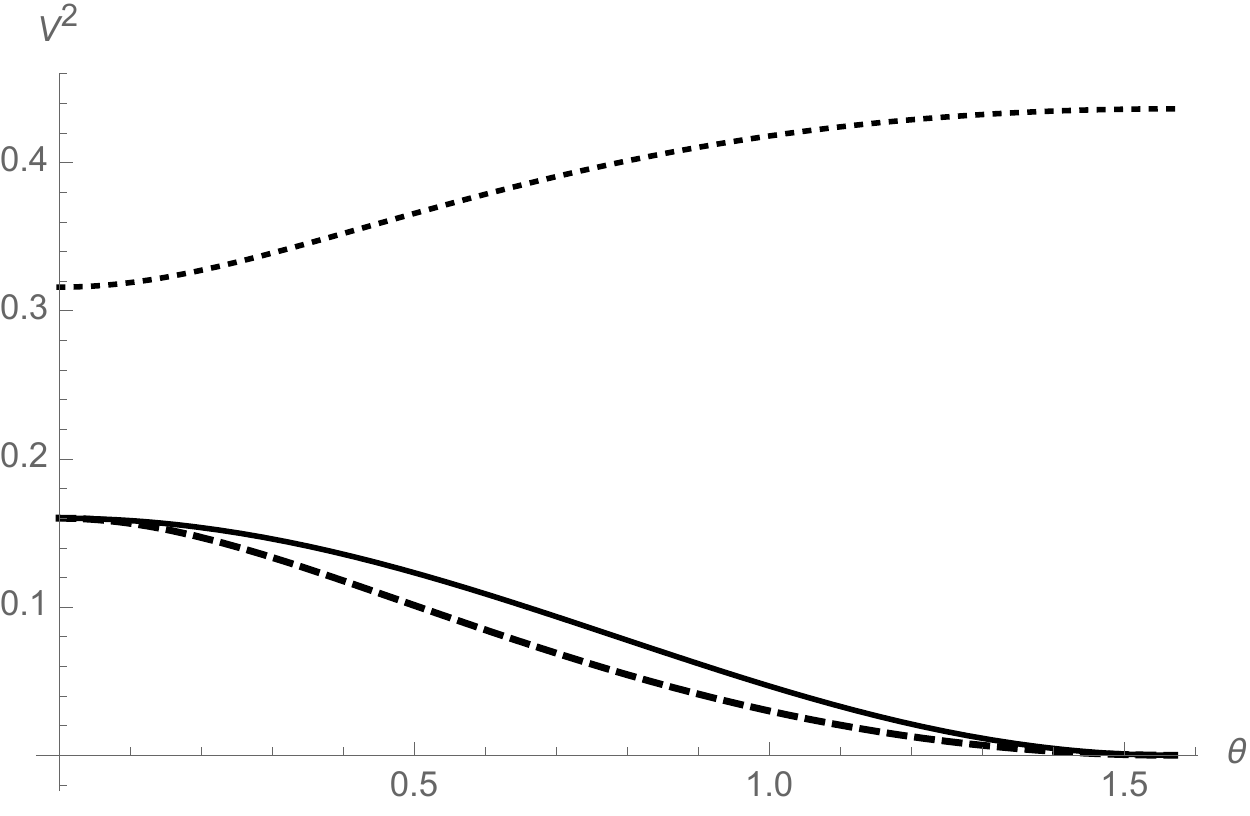}
\caption{Velocities from high temperature equation of state \eqref{pweak} with $\mu/T^2 = 0.4$, $a=g=\beta=1$.}
\label{fig:highT}
\end{subfigure}
\begin{subfigure}{0.4\linewidth}
\centering
\includegraphics[width=\linewidth]{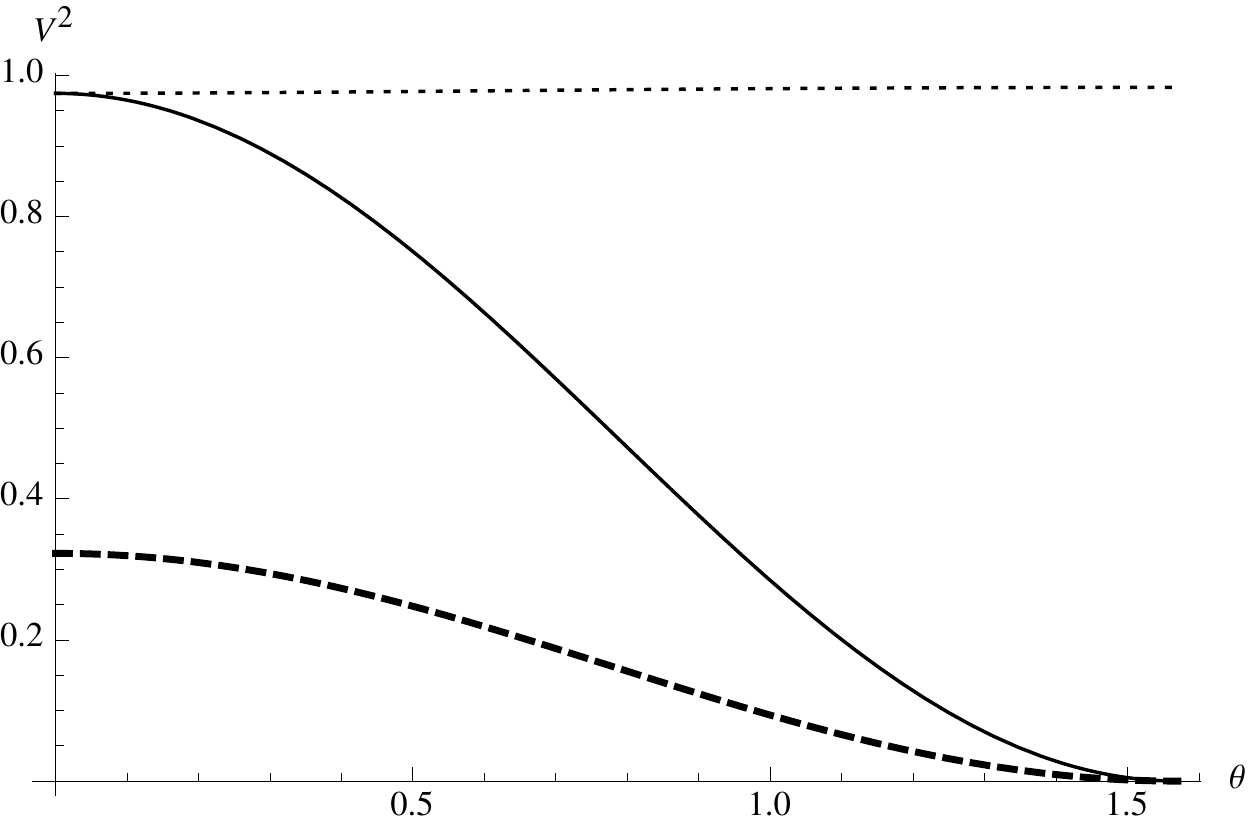}
\caption{Velocities from high field equation of state \eqref{pstrong} with $T^2/\mu = 0.4$, $a'=g'=\beta'=1$.}
\label{fig:lowT}
\end{subfigure}
\caption{Sample plots illustrating velocity squared of Alfven wave $v_A^2$ (solid black line), fast magnetosonic wave $(v_M^2)_{\mathrm{fast}}$ (dotted line), and slow magnetosonic wave $(v_M^2)_{\mathrm{slow}}$ (dashed line) as a function of angle $\theta$ between the momentum of the wave and the background magnetic field.}
\label{fig:PhVel}
\end{figure*}

We present some representative plots of the velocities in the strong-field expansion in Figure \ref{fig:lowT}. 

\subsection{Cyclotron mode}

Lastly, let us mention that by introducing a non-trivial curvature for the source of our generalized charge, we can recover the familiar cyclotron mode of plasma physics in the presence of a finite electric charge. At zero spatial momentum in the presence of a spatial and isotropic external field $H_{ijk} = n \epsilon_{ijk}$, we find that the system can undergo cyclotron motion with frequency
\begin{align}
\omega = \pm \frac{2 n}{ \mu \sqrt{ 1 + \frac{Ts}{\mu\rho} } }  .
\end{align}

\section{Magnetohydrodynamics with a baryon number} \label{sec:baryon}
The theory that we have developed above demonstrates the essential physics in the hydrodynamics of conserved flux tubes. However, for phenomenological applications we should extend it slightly to also include a conventional $0$-form global symmetry (e.g. baryon number). This turns out to be entirely straightforward, and thus we present here only results without derivations. 

We denote the baryon number current by $B^{\mu}$. We have $\nabla_{\mu} B^{\mu} = 0$, and we denote its conserved charge and chemical potential by $n_B$ and $\mu_B$. The thermodynamic relations of interest are:
\begin{align}
&\ep + p = \mu_B n_B + \mu \rho + T s, \\
&dp = s \,dT + \rho\, d\mu + n_B \,d \mu_B.
\end{align}
At the level of ideal hydrodynamics, the relations \eqref{idealTJ} remain unaltered, and the expression for the baryon current is
\be
B^{\mu}_{(0)} = n_B u^{\mu}.
\ee
The conservation equation for $T^{\mu\nu}$ is modified to include a contribution from an external gauge field $F_B = dA_B$ that couples to the baryon current:
\be
\nabla_{\mu} T^{\mu\nu} = H^{\nu}_{\phantom{\nu}\rho\sig}J^{\rho\sig} + F_B^{\nu\mu} B_{\mu} . 
\ee
Just as before, the ideal hydro equations result in a conserved entropy current $s u^{\mu}$ at the ideal hydrodynamic level. 

If we move to first order hydro, the canonical form for the entropy current is
\be
S^{\mu} = s \,u^{\mu} - \frac{1}{T}T^{\mu\nu}_{(1)}u_{\nu}  - \frac{\mu}{T} J^{\mu\nu}_{(1)}h_{\nu} - \frac{\mu}{T} B^{\mu}_{(1)}.
\ee
We expand the first order correction to the baryon current as
\be
B^{\mu}_{(1)} = \delta n_B u^{\mu} + q h^{\mu} +f^{\mu} ,
\ee 
with $\delta n_B$ and $q$ first-order scalars and $f^{\mu}$ a transverse vector. It is convenient to use scalar redefinitions of the baryon chemical potential to set $\delta n_B = 0$; then an entropy analysis similar to that above results in the following expressions:
\begin{align}
q & = -\sig_{\parallel} h^{\mu}\le(T \p_{\mu}\le(\frac{\mu_B}{T}\ri) - F^B_{\mu\nu} u^{\nu}\ri) , \\
f^{\mu} & = -\sigma_{\perp} \Delta^{\mu\nu} \le(T \p_{\nu}\le(\frac{\mu_B}{T}\ri) - F^B_{\nu\rho} u^{\rho}\ri) ,
\end{align}
where the transport coefficients $\sigma_{\perp,\parallel}$ are simply conventional global (if anisotropic) conductivities for the baryon current. 

\section{Relativistic MHD at zero temperature} \label{sec:T0}

In this section, we turn our attention to MHD at zero temperature. What do we mean by this? Normally, a system at $T=0$ is outside the hydrodynamic limit as there is no way to properly define a long wave-length limit compared with typical decay widths. Another way of stating this fact is the following: at $T=0$ there is no dissipation and this leads to long-lived excitations that do not arise necessarily as a consequence of conservation laws, leading to the presence of gapless modes that need to be included in the description of these systems. Also common is the pressence of gapless Goldstone modes when there is a broken symmetry.

There is however a physical situation of interest that can be described from a hydrodynamic perspective. Consider a system at a very low temperature compared with the scale set by $\mu$. Is there a universal description of this system? The answer is typically no. The reason is that there are finite energy excitations around this equilibrium state that require understanding of the finite temperature theory. An example of this feature is given by the physics of the slow magnetosonic mode described in section 
\ref{Tlessmu}. The situation is even more severe if more light modes need to be included as $T \rightarrow 0$. One can, instead, consider a subsector of the theory where temperature fluctuations are not allowed. In this case, we are effectively studying a small part of the Hilbert space at energies $E \sim \mu Q$. In order to do this we can then restrict to $T=0$ in our equation of state and study the universal properties implied by the conservation of charge. Notice that an important feature of this system is that entropy is typically zero (i.e. the number of states we are considering is not exponentially large) according to the third law of thermodynamics. A curious counter-example to this seem to be holographic systems at finite density for standard conserved 1-form currents \cite{Chamblin:1999tk}. We will be agnostic about this for now and come back to this issue at the end of this section around equation (\ref{divform}).

The program above can be carried out in standard hydrodynamics of conserved 1-form currents. See Appendix \ref{hydro0} for a short discussion. The physics of this system is, however, conceptually not fundamentally different from usual hydrodynamics at $T \neq 0$. The reason is that the symmetry breaking pattern of a theory at finite density but $T=0$ is identical to the $T \neq 0$ situation. It is always the case that the $SO(3,1)$ Lorentz symmetry is broken down to $SO(3)$. This is because the charge density still selects a rest frame, even at $T=0$. 

The situation is much more interesting in our description of magnetohydrodynamics. At $T \neq 0$ and $\mu=0$ we have the usual symmetry breaking pattern $SO(3,1) \rightarrow SO(3)$.  At $\mu \neq 0$, more symmetries are broken: the presence of background magnetic fields leads to the choice of a preferred spatial direction, further breaking Lorentz symmetry as $SO(3,1) \rightarrow SO(2)$ and leading to the richer theory described in the previous sections. What happens at $T = 0$ and $\mu \neq 0$? Interestingly, the situation is different from the case discussed before. At $T=0$, there is only an anti-symmetric tensor turned on in the background responsible for the magnetic field. This configuration is invariant under Lorentz boosts along the magnetic field lines. Therefore, in this case we have an enhanced symmetry $SO(3,1) \rightarrow SO(1,1) \times SO(2)$ with respect to the finite temperature case. 

This novel symmetry breaking pattern implies that the thermodynamic variables necessary to describe the system are completely different from the discussion in Section \ref{sec:ideal}.  In what follows we describe this new theory. There are concrete applications of this formalism in the understanding of systems at strong magnetic fields, such as some astrophysical systems.

\subsection{Effective degrees of freedom}

The relevant hydrodynamic fields are a scalar chemical potential $\mu$ and an antisymmetric $u_{\mu\nu}$ field that parameterizes the rest frame enjoying $SO(1,1) \times SO(2)$ symmetry. We normalize it as
\begin{align}\label{uNorm}
u_{\mu\nu} u^{\mu\nu} = - 2 .
\end{align}
As in usual hydrodynamics, the normalization is possible as $\mu$ carries this information. What is crucial, however, is the sign above, signalling that the tensor above is ``mostly" in the plane acted on by $SO(1,1)$. This is similar to the familiar $u_\mu u^\mu = -1$ constraint.
We would like, however, $u^{\mu\nu}$ to satisfy a stronger constraint and live entirely in the plane acted on by the $SO(1,1)$, so there exists a frame where the charge is entirely at rest. This is enforced in a covariant manner by further demanding that
\begin{align}\label{uNorm2}
u^{\mu\nu} u_{\nu\rho} u^{\rho\sigma} = u^{\mu\sigma}.
\end{align}
It will be convenient to introduce a symmetric tensor 
\begin{align}
\Omega^{\mu\nu} \equiv u^{\mu\lambda} u_\lambda^{\phantom{\lambda} \nu}.
\end{align}
The tensor $\Om$ is the $SO(1,1)$-invariant metric on the 2d subspace spanned by the magnetic field. It projects any index onto this subspace. We will also make use of the projector orthogonal to this subspace, which projects onto the $SO(2)$-invariant sector:
\begin{align}
\Pi^{\mu\nu} & \equiv g^{\mu\nu} - \Om^{\mu\nu},
\end{align}
with trace $\Pi^{\mu}_{~\mu} = \Om^{\mu}_{~\mu} = 2$. Henceforth, we will focus on the theory in flat space, $g_{\mu\nu} = \eta_{\mu\nu}$.  It will also be useful to visualise our construction in a Cartesian coordinate system aligned to the magnetic field by setting $u^{tz} = - u^{zt}=1$, while all remaining components are zero. The $SO(1,1)$ group then acts on $(t,z)$ and leaves invariant the metric $\Om_{\mu\nu}$. $SO(2)$ acts on $(x,y)$ and leaves invariant the metric $\Pi_{\mu\nu}$. 

We can now write down the most general stress-energy tensor and the antisymmetric conserved $2$-form in flat space:
\begin{equation}\label{idealTJzeroT}
\begin{aligned}
T^{\mu\nu}_{(0)} & = -\ep \, \Omega^{\mu\nu}  + p \, \Pi^{\mu\nu},  \\
J^{\mu\nu}_{(0)} & = \rho \, u^{\mu\nu} ,
\end{aligned}
\end{equation}
where $\ep$, $p$ and $\rho$ are functions of $\mu$ only.  It is also important to note that the thermodynamic relation \eqref{thermo} has now become
\begin{align}
\ep + p = \mu\rho, && dp = \rho \,d\mu . \label{thermoZeroT}
\end{align}

We can recover the zero temperature hydrodynamic theory of \eqref{idealTJzeroT} from the finite temperature theory \eqref{idealTJ} by the following identification of the hydrodynamic variables:
\begin{align}
u^{\mu\nu} = 2 \, u^{[\mu} h^{\nu]},
\end{align}
keeping $\mu$ finite and sending $T$ to zero. In this language, the symmetric $SO(1,1)$ and $SO(2)$ metrics are
\begin{align}
\Omega^{\mu\nu} &= h^{\mu}h^\nu - u^\mu u^\nu, \\
\Pi^{\mu\nu} &= g^{\mu\nu}  + u^\mu u^\nu - h^{\mu}h^\nu = \Delta^{\mu\nu} ,
\end{align}
where $\Delta^{\mu\nu}$ is the finite temperature projector from Eq. \eqref{FiniteTProjector}. 

At $T^2\ll \mu$, we, therefore, expect a universal subsector of hydrodynamics to satisfy, in the ideal limit, the dynamical conservation equations

\bea
\nabla_\mu T_{(0)}^{\mu \nu} &=& 0\, , \\
\nabla_\mu J_{(0)}^{\mu \nu} &=& 0\, ,
\eea
\noindent with the constituent relations (\ref{idealTJzeroT}).

As always, it is important to check that this system of equations closes and is not overdetermined. The system we are considering has a priori 5 degrees of freedom given by the scalar $\mu$ and the four degrees of freedom in $u_{\mu \nu}$ subject to the constraints (\ref{uNorm}) and (\ref{uNorm2}). These four degrees of freedom can be viewed, in terms of symmetries, as the non-trivial action of the Lorentz group on a tensor preserving the $SO(1,1) \times SO(2)$ symmetries. 

On the other hand, we see that the system above consists, naively, of eight equations of motion. This presents a danger, as the system of equations could be overdetermined. This is, however, not so. One of the equations (the time component of the charge conservation) is actually a constraint, as in the $T \neq 0$ case. This constraint is consistently propagated by the other equations of motion. Thus, enforcing this constraint removes two equations of motion and one degree of freedom, still leaving an excess of two equations. For this system to not be overdetermined they need to be trivial. Luckily this is exactly the case for (\ref{idealTJzeroT}). Consider the equation

\be\label{trivialeq}
\left(\nabla_\mu T^{\mu\nu}\right) \Omega_{\nu \lambda} + \mu \left(\nabla_\mu J^{\mu \nu}\right) u_{\nu \lambda} =0  \, .
\ee

This equality is satisfied off-shell for any field $u_{\mu \nu}$ satisfying the constraints (\ref{uNorm}) and (\ref{uNorm2}), and thus the two equations of motion that it contains are redundant. Therefore, the system of equations under consideration is consistent as a full set of non-linear partial differential equations. Interestingly, as we elaborate on in Appendix \ref{hydro0}, (\ref{trivialeq}) can be viewed as the natural zero-temperature generalization of the equation for the conservation of the entropy current at finite temperature. 

\subsection{Beyond ideal hydrodynamics}

We now move beyond ideal hydrodynamics. From the structure of available tensors, it is clear that there are no suitable one-derivative structures as they would have an odd number of indices. Therefore $T^{\mu\nu}_{(1)} = J^{\mu\nu}_{(1)} = 0$. The leading order corrections only enter at the level of second-order hydrodynamics. This observation, which follows only from available tensor structures is consistent with the fact that at $T=0$, the theory is expected to be non-dissipative. Since first-order corrections to ideal hydrodynamics are normally purely dissipative, such corrections should be absent. 

To determine the form of the potential second-order corrections, we first discuss the decomposition of an arbitrary tensor under $SO(2) \times SO(1,1)$. Using $a, b, \ldots$ for $SO(1,1)$ indices and $i, j, \ldots$ for $SO(2)$ indices, an antisymmetric tensor $s^{\mu\nu}$ breaks into three blocks: $s^{ab}$ and $s^{ij}$, which are tensors under $SO(1,1)$ and $SO(2)$ respectively, as well as the off-diagonal elements $s^{ia}$ that transform as a product of vectors under $SO(1,1)$ and $SO(2)$ (denoted as $v \otimes v$). A similar classification holds for symmetric tensors $t^{\mu\nu}$, except that we can also extract out the scalar traces of the tensors $t^{ab}$ and $t^{ij}$. 

The three projectors onto the tensor representation of $SO(1,1)$, $SO(2)$, and vector representations of $SO(1,1) \otimes SO(2)$ are 
\begin{align}
SO(1,1): && P^{\mu\nu}_{(\omega)\rho\sig} &= \Omega^{\mu}_{~\rho} \Omega^{\nu}_{~\sig}, \\
SO(2): && P^{\mu\nu}_{(\Pi)\rho\sig} &=  \Pi^{\mu}_{~\rho} \Pi^{\nu}_{~\sig} , \\
v \otimes v: && P^{\mu\nu}_{(v)\rho\sig}  &= \Om^{\mu}_{~\rho} \Pi^{\nu}_{~\sig} + \Pi^{\mu}_{~\rho} \Om^{\nu}_{~\sig},\label{Projectorvv}
\end{align}
with traces $P^{\mu}_{(\omega)\mu\rho\sig} =\Omega_{\rho\sigma}$, $P^{\mu}_{(\Pi)\mu\rho\sig} =  \Pi_{\rho\sigma}$ and $P^{\mu}_{(v)\mu\rho\sig}  = 0$. Hence, the symmetric and traceless projectors of an arbitrary matrix $M^{\mu\nu}$ onto the three sectors are
\begin{align}
SO(1,1): &&   \left( P^{(\mu\nu)}_{(\omega)\rho\sig} - \frac{1}{2} \Omega^{\mu\nu} \Omega_{\rho\sigma} \right) &M^{\rho\sigma}, \label{ProjectorsFinal1} \\
SO(2): && \left(P^{(\mu\nu)}_{(\Pi)\rho\sig}  - \frac{1}{2} \Pi^{\mu\nu} \Pi_{\rho\sigma} \right) &M^{\rho\sigma}, \label{ProjectorsFinal2}\\
v \otimes v: && P^{(\mu\nu)}_{(v)\rho\sig} & M^{\rho\sigma},\label{ProjectorsFinal3}
\end{align}
and the antisymmetric parts follow from ${P^{[\mu\nu]}}_{\rho\sig}  M^{\rho\sigma}$ in all three cases.

We now use this classification to parametrize the most general correction to \eqref{idealTJzeroT}, i.e. the analog of \eqref{1stordercorr} at zero temperature:
\begin{equation}
\begin{aligned}
T^{\mu\nu}_{(2)}  =& -\delta \ep\;\Om^{\mu\nu} + \delta p\;\Pi^{\mu\nu} \\
& + t_{SO(1,1)}^{\mu\nu} + t_{SO(2)}^{\mu\nu} +t_{v \otimes v}^{\mu\nu} , \\
J^{\mu\nu}_{(2)} =& \,\, \delta \rho\;u^{\mu\nu} + s^{\mu\nu}_{SO(2)} + s_{v \otimes v}^{\mu\nu}. \label{JT0corr}
\end{aligned}
\end{equation} 
Here, a two-index object with the $SO(2)$ or the $SO(1,1)$ subscript indicates that the object transforms as a tensor under the appropriate group, whereas the $v \otimes v$ subscript indicates that it transforms as a product of vectors under $SO(2)$ and $SO(1,1)$. Note that in the antisymmetric sector, any putative $s^{\mu\nu}_{SO(1,1)}$ is proportional to $u^{\mu\nu}$, and thus any corrections from those terms have been included in the $\delta \rho\;u^{\mu\nu}$ term. 

We now note, paralleling the discussion around \eqref{1stordercorr}, that we may exploit the possibility to change the hydrodynamic frame under a scalar redefinition $\mu \to \mu + \delta \mu$ and an antisymmetric tensor redefinition $u^{\mu\nu} \to u^{\mu\nu} + \delta u^{\mu\nu}$. The redefinitions are subject to two constraints: firstly, the perturbation of the normalization constraint \eqref{uNorm} results in 
\be
u_{\mu\nu} \delta u^{\mu\nu} = 0  . \label{traceU}
\ee
Secondly, the perturbation of the subspace constraint \eqref{uNorm2} can be written in the following form:
\begin{align}\label{subU}
\le(P^{\mu\nu}_{(\Pi)\rho\sig} - 2 P^{\mu\nu}_{(\om)\rho\sig}  \ri)\delta u^{\rho\sig} = 0 ,
\end{align}
This annihilates the part of $\delta u^{\mu\nu}$ living in the $SO(1,1)$ and $SO(2)$ tensor blocks, meaning that it lives entirely in the (antisymmetric part of the) vector $\otimes$ vector block. This automatically implies the first constraint, i.e. Eq. \eqref{traceU}. 


By using the scalar redefinitions of $\mu$, we may eliminate a single scalar correction in \eqref{JT0corr}, which we take to be $\delta\rho$. Furthermore, using the antisymmetric tensor redefinition of $u^{\mu\nu}$, we may further remove the vector $\otimes$ vector term $s_{v \otimes v}^{\mu\nu}$. Thus, we find the simpler frame-fixed structure of corrections to the conserved hydrodynamic tensors to be
\begin{equation}
\begin{aligned}
T^{\mu\nu}_{(2)} =& -\delta \ep\;\Om^{\mu\nu} + \delta p\;\Pi^{\mu\nu} \\
& + t_{SO(1,1)}^{\mu\nu} + t_{SO(2)}^{\mu\nu} +t_{v \otimes v}^{\mu\nu} , \\
J^{\mu\nu}_{(2)} =&\,\,  s^{\mu\nu}_{SO(2)}   . 
\end{aligned} 
\end{equation}
These expressions are analogous to the finite temperature expressions in Eq. \eqref{corr}. The task is now to determine these six corrections in terms of the available second derivative objects. For the purposes of this paper, we will only focus on terms that play a role in linearized hydrodynamics and enter into linear dispersion relations. As is well known in the hydrodynamics literature \cite{Baier:2007ix,Bhattacharyya:2012nq,Grozdanov:2015kqa}, a full nonlinear analysis even at a two derivative order typically involves tens of possible structures.

We expand the effective degrees of freedom to linear order,
\begin{align}\label{linear}
\mu \rightarrow \mu + \delta \mu, && u_{\mu \nu} \rightarrow u_{\mu \nu} + \delta u_{\mu \nu},
\end{align}
\noindent where we take $\mu$ and $u_{\mu \nu}$ to be constant. Note that to linear order we can write all structures efficiently in momentum space spanned by $k^\mu$. Furthermore, it will come in handy to make use of the decomposition into $SO(1,1)$ and $SO(2)$ indices. Following the notation above
\bea
u_{\mu \nu} & \rightarrow & u_{a b} \, ,\\
\delta u_{\mu \nu} & \rightarrow & \delta u_{\left[ a i \right]}  \, , \\
k^\mu  & \rightarrow & \left( \omega^a, q^i \right)  .
\eea
We find in this approximation the following allowed tensor structures\footnote{Here we have demanded that corrections to the energy momentum tensor should be even under $u_{\mu \nu} \rightarrow - u_{\mu \nu}$, $\delta u^{a i} \rightarrow - \delta u^{a i}$, as the ideal term. This is consistent with the charge assignments in Table \ref{tbl:disc}.}:
\bea
\delta \varepsilon &=& \zeta_{(1,1)} \, \left[  q_i \, \delta u^{a i} \, u_{a b} \,  \omega^b \right]  ,\label{s1}\\
\delta p &=& \zeta_{(2)} \, \left[  q_i \, \delta u^{a i} \, u_{a b} \,  \omega^b \right]  ,\label{s2}\\
t^{a b}_{SO(1,1)} &=&   \eta_{(1,1)} \, \left[2 q_i \, \delta u^{c i} \, u_{c}^{\phantom{c} (a} \,  \omega^{b)} \right.\nonumber \\
  & & \left.   -   \Omega^{a b}  q_i \, \delta u^{c i} \, u_{c d} \,  \omega^d \right]  ,\label{s3}\\
t^{i j}_{SO(2)} &=&  \eta_{(2)} \, \left[ 2 q^{(i} \, \delta u^{a  j) } \,  u_{a b} \,  \omega^b \right. \nonumber \\
  & & \left.   -   \Pi^{i j}  q_k \, \delta u^{a k} \, u_{a b} \,  \omega^b \right]  ,\label{s4}\\
t_{v \otimes v}^{a i}  &= & t_{v \otimes v}^{i a}  =  \nu_0 \left[ q_j \delta u^{b j} u_{b}^{\phantom{b} a} q^i  \right] \nonumber\\
& & + \nu_1 \left[ \delta u^{b i} \, u_{b}^{\phantom{b} a} \omega^c \omega_c\right] + \nu_2 \left[ \delta u^{b i} \, u_{b}^{\phantom{b} a} q^j q_j\right] , \quad\quad\, \label{s5}\\
s^{i j}_{SO(2)} &=& 0 \, , \label{s6}
\eea
\noindent with transport coefficients parameterized by $ \zeta_{(1,1)}$, $\zeta_{(2)}$, $\eta_{(1,1)}$, $\eta_{(2)}$, $\nu_0$, $\nu_1$, $\nu_2$. Note that at linear order no antisymmetric tensors in the $SO(2)$ sector remain after using the equations of motion of the ideal system. The details behind this construction are described in Appendix \ref{app:T0Tensors}. 

Importantly, it turns out that not all of these corrections are permitted for a consistent system: we discuss this in the next section.

\subsection{Application: Modes of the system}

To find the modes of MHD at zero temperature, we apply the same procedure as in Section \ref{sec:ModesFiniteT}.  We  choose coordinates such that the background is given by $u^{tz} = - u^{zt} = 1$ and the other components equal to zero. In our covariant notation from before, we identify $t,z$ with $a, b, \ldots$ and $x, y$ with $i, j, \ldots$.

 As in the previous case, we can always use the $SO(2)$ symmetry to rotate the $q^i$ momentum into the $x$ axis. But in this case we can also make use of the $SO(1,1)$ symmetry acting on the $(t,z)$ plane. Differently from the $SO(2)$ case, here, there are two different representations of $SO(1,1)$ that we can consider\footnote{In principle, one could also have spacelike representations $\omega^a \omega_a >0$. They correspond, however, to dissipative modes not present in this non-dissipative $T=0$ theory.}. An $SO(1,1)$ vector $\omega^a$ can be:
 
\begin{itemize}
\item lightlike: ~~$\omega^a \omega_a = 0$,
\item timelike: ~~$\omega^a \omega_a <0$.
\end{itemize} 

Let us first consider the lightlike case and choose $\omega^t = \omega$, $\omega^z = \omega$, $q^x = q$, $q^y=0$. Solving the system of equations, we find this choice is a solution for excitations in the Alfv\'{e}n channel,

\be
\delta u^{i a} q_i = \delta u^{i a} \omega_a=0 \, , \quad  \delta\mu=0 \, .
\ee
Notice that this implies that Alfv\'{e}n waves have dispersion relations manifestly independent of $q$. Using the familiar representation familiar from the finite temperature case,
\be
k^\mu = (\omega ,q ,0,k) \equiv (\omega, \kappa \sin\theta,0,\kappa\cos\theta),
\ee
\noindent we obtain the dispersion relation
\be\label{alft0}
\quad \omega_{A} = k = \kappa \cos \theta \, .
\ee
This dispersion relation is also manifestly independent of all transport coefficients. One might have expected that higher derivative corrections could have modified it, making the Alfv\'{e}n mode into a timelike representation of $SO(1,1)$. This is not the case, however. Furthermore, it is simple to see that this situation persists to all orders in derivatives. A correction to the linear equations of motion appears in the form of a vector linear in $\delta u^{i a}$. In order to make a vector, one of these two indices needs to be contracted, but for Alfv\'{e}n waves all contractions vanish and the dispersion relation cannot be corrected at any order in derivatives. The result (\ref{alft0}) is, therefore, completely universal in any magnetohydrodynamical theory at high enough $\mu \gg T^2$ and it agrees with our large $\mu$ expansion (\ref{muexp1}). The mode corresponds to the transverse fluctuation of magnetic field lines. This fluctuation can only move in the direction of the magnetic field and at the speed of light as seen from (\ref{alft0}).

Now, consider the second, timelike case. It is now possible to use the symmetries to choose $\omega^t = \omega$, $\omega^z = 0$, $q^x = q$, $q^y=0$. These modes correspond to magnetosonic excitations. In this case the higher derivative corrections do contribute to the dispersion relation.

There remains an important point that constrains these higher derivative corrections. In the previous discussion concerning the ideal fluid at $T=0$, it was of crucial importance that two equations in the system \eqref{trivialeq} happened to be trivial for the system not to be overdetermined. This property of the differential equations is lost once the higher derivative corrections are included. It appears that ensuring the system to not be overdetermined is a more dramatic version of the usual requirements that apply to the divergence of the entropy current (in standard hydrodynamics) and severely restricts the possible number (and signs) of transport coefficients. 

A complete resolution of the issue and enumeration of the resulting constraints on second order coefficients would demand understanding the full non-linear behavior of the theory, which lies outside the scope of this article. For linear perturbations, it is interesting to remark that the formerly trivial equation (\ref{trivialeq}) can be written as a total divergence:
\bea
\left(\nabla_\mu T^{\mu\nu}\right) \Omega_{\nu \lambda} + \mu \left(\nabla_\mu J^{\mu \nu}\right) u_{\nu \lambda} \sim \nonumber\\
 \nabla_\mu \left( T^{\mu\nu} \Omega_{\nu \lambda} + \mu J^{\mu \nu} u_{\nu \lambda} - p \, \Omega^\mu_{\phantom{\mu} \lambda} \right)   \label{divform} \, .
\eea
Therefore, a sufficient but not necessary condition for consistency at the linear level is to demand that the term inside the divergence vanish identically. This is somewhat reminiscent of a third law of thermodynamics at $T=0$. This condition would have the effect of setting  $ \zeta_{(1,1)}=\eta_{(1,1)}=\nu_0=\nu_1=\nu_2=0$, leaving only $\zeta_{(2)}$ and $\eta_{(2)}$ available. Unfortunately, this condition is not sufficient at the non-linear level; terms that have been omitted from \eqref{divform} can spoil the consistency relation. 

Therefore in this article we take a more conservative stance and simply demand the minimal condition for the {\it linear} system to be self-consistent. Explicit solution of the equations of motion for the magnetosonic mode yields the necessary condition:
\be
\nu_0 = -\nu_2 + \frac{\rho}{\mu \chi} \left(\nu_1 - \eta_{(1,1)} + \zeta_{(1,1)} \right) ,
\ee
\noindent and gives the following dispersion relation:
\begin{align}
\omega_M &= \pm \sqrt{\frac{\rho}{ \mu \chi}} \, q \,\, \bigg{(}1  \nonumber\\
 & \left.+ \frac{1}{2 \mu^2 \chi} \left[\left(\zeta_{(1,1)} - \eta_{(1,1)}\right) - \frac{\mu \chi}{\rho} \left(\zeta_{(2)} + \eta_{(2)} \right) \right] q^2 \right)\, ,\quad\quad\quad
\end{align}

\noindent where $\chi = \frac{\partial \rho}{\partial \mu}$. Boosting this result in the $(t,z)$ plane, we get
\be
\omega_M =  \pm \left(v_M \kappa + \alpha_M \kappa^3\right),
\ee
\noindent with
\be
v_M = \sqrt{ \cos^2 \theta + \frac{\rho}{ \mu \chi} \sin^2 \theta} 
\ee
and
\begin{align}
\alpha_M =&\,\, \frac{\rho}{ 2\mu^3 \chi^2} \frac{ \sin^4 \theta}{ \sqrt{ \cos^2 \theta + \frac{\rho}{ \mu \chi} \sin^2 \theta}}  \nonumber\\
 & \times \left[\left(\zeta_{(1,1)} - \eta_{(1,1)}\right) - \frac{\mu \chi}{\rho} \left(\zeta_{(2)} + \eta_{(2)} \right) \right]    .
\end{align}

In the particular case when no other scale enters the problem, we must have $\frac{\rho}{ \mu \chi}=1$ and the formulae above agree with (\ref{muexp2}). In this case, the speed of the universal magnetosonic mode is that of light in the ideal limit, but it can be corrected at higher order in the derivative expansion. As expected, this must impose causality constraints on the transport coefficients.

\section{Conclusions and outlook} \label{sec:conc}
In this work, we have systematically developed the hydrodynamic theory of a 2-form current at finite temperature, describing the dynamics of a finite density of strings as it relaxes towards thermal equilibrium. As we have emphasized, the conventional theory of relativistic magnetohydrodynamics fits into this general class of theories, where the density of strings represents magnetic flux. Thus our work may be viewed as a construction of a generalized theory of MHD from the point of view of symmetries and effective field theory, making no reference to any sort of weak electromagnetic coupling. For a particular choice of equation of state (one in which temperature and chemical potential decouple entirely) our theory reduces to conventional MHD, but this choice is not required on effective field theory grounds. 

We also worked to first order beyond ideal hydrodynamics, finding that there are a total of seven transport coefficients that can be interpreted as anisotropic resistivities and viscosities. Along the way, we provide a precise definition of a resistivity for a dynamical $U(1)$ gauge theory and explain how this resistivity can be computed microscopically from Kubo formulas involving correlation functions of the electric field operator. It is interesting to note that we expect that our universal resistivity will precisely coincide with the inverse of the conventionally defined conductivity of the ``gauged''  electric current only at weak electromagnetic coupling.

While our theory is conceptually satisfying, it also makes precise physical predictions. For example, as a first step, we studied small fluctuations around thermal equilibrium, identifying within our framework well-known magnetohydrodynamic modes. We systematically study of the dissipation of these modes, obtaining (to the best of our knowledge) novel physics, such as angle-dependent dissipation of Alfv\'en and magnetosonic waves. In principle, these predictions are open to experimental verification. 

We then turned to a truncation of the theory at low temperatures: this involves an emergent Lorentz symmetry along the background magnetic field lines, and as a result hydrodynamic fluctuations involve a fluid tensor rather than a fluid velocity. This leads to an interesting generalization of hydrodynamics. We initiated the study of this new framework, working out the structure of linear perturbations and listing possible higher derivative corrections to second order in fluid momenta. We find it particularly interesting that a consistency condition on the closure of the differential equations appears to play a role analogous to the constraints imposed by the second law of thermodynamics in conventional hydrodynamics. This deserves further exploration.

There are many directions for future research. One could imagine studying this hydrodynamic theory from holography, by developing the magnetohydrodynamic analog of the fluid-gravity correspondence \cite{Bhattacharyya:2008jc}. The dual of the $2$-form current will now be a $2$-form gauge field propagating in a five-dimensional bulk. It turns out that the background black hole solution corresponding to an equilibrium fluid is S-dual to the black holes studied in \cite{D'Hoker:2009mm}, and our theory could be tested by studying perturbations around this background. 

A further direction is to understand the formulation of MHD, in light of our work, in the language of the effective field theory of Goldstone bosons for space-time symmetry breaking along the lines of \cite{Nicolis:2013lma}. As generalized global symmetries provide new examples with potentially novel symmetry breaking patterns, these theories provide a natural arena to test the power of the formulation in \cite{Nicolis:2013lma}. Such a framework may allow for the systematic study of fluctuations. It would also be interesting to go to higher orders in hydrodynamics and to understand questions regarding the stability of MHD in our formalism\footnote{See \cite{Torrieri:2016dko} for a recent discussion of stability of (Navier-Stokes) hydrodynamics in connection with the Israel-Stewart theory.}. 

Finally, we stress that our theory differs from that of traditional MHD, as it separates the universal constraints of symmetry and the effective theory of hydrodynamics from non-universal artefacts that arise from assuming a weak electromagnetic coupling. As we explain in the introduction, the effective strength of electromagnetic interactions in sufficiently dense plasmas can be very large indeed, and it would be very interesting to explore if our generalized theory of strings could be experimentally relevant for plasma physics.

\vspace{0.2in}   \centerline{\bf{Acknowledgements}} \vspace{0.2in} We thank A.~Belin, J.~Bhattacharya, S.~Hartnoll, P.~Kovtun, H.~Liu, J.~McGreevy, S.~Minwalla, A.~Nicolis, N.~Poovuttikul, D.~T.~Son, A.~Starinets and G.~Torrieri for illuminating discussions. This work is part of the Delta ITP consortium, a program of the NWO that is funded by the Dutch Ministry of Education, Culture and Science (OCW). 

\begin{appendix}

\section{Ideal fluid partition function and thermodynamics} \label{app:partition}

In this Appendix, we follow \cite{Banerjee:2012iz,Jensen:2012jh} to derive the equilibrium form of the ideal stress-energy tensor and current \eqref{idealTJ}
\begin{align}
T^{\mu\nu}_{(0)} & = (\ep + p)\, u^{\mu}u^{\nu} + p \, g^{\mu\nu} - \mu\rho\, h^{\mu}h^{\nu}, \nonumber \\
J^{\mu\nu}_{(0)} & = 2\rho \, u^{[\mu}h^{\nu]} , \label{TJapp}
\end{align}
from a partition function. This is done by writing down the form of the most general Euclidean partition function in the presence of static (but spatially dependent) metric and $2$-form gauge field sources. Gauge and diffeomorphism invariance constrain the form of this partition function, providing an efficient way to derive constraints the hydrodynamic stress-energy tensor and currents. As we are performing a static Euclidean computation this only allows access to objects that can be probed by {\it equilibrium} fluid flows. 

We should first define what is meant by ``equilibrium''. In conventional fluid dynamics, to have an equilibrium solution one requires a timelike Killing vector $\xi$ to define the direction of fluid flow $u^{\mu} \sim \xi^{\mu}$. We will also require an additional spacelike Killing vector $\zeta$ to define the direction of the strings $h^{\mu} \sim \zeta^{\mu}$. Note that with no loss of generality we may take $\xi \cdot \zeta = 0$. Provided that these two vectors have vanishing Lie bracket, $\sL_{\xi} \zeta = 0$, it is straightforward to show that such a field profile satisfies the equations of motion arising from the conservation of \eqref{TJapp}, corresponding to a fluid at rest with the strings oriented in the direction of $\zeta$. 

As $\xi$ and $\zeta$ have vanishing Lie bracket, we can take them to define a coordinate system $(t,z)$. Thus we may write the background Lorentzian background metric of the form
\begin{align}
ds^2 & = g_{ij} \left(x^k\right)dx^i dx^j \nonumber \\ 
& + g_{ab}\left(x^i\right) \left(dx^a - A^{a}_i  \left(x^k\right) dx^i\right)\left(dx^b - A^{b}_i \left(x^k\right) dx^i\right) . \label{lor}
\end{align}
Where here $a,b$ run over $t,z$ and $i,j$ run over the remaining coordinates. Furthermore, since $\xi \cdot \zeta = 0$, we may take the metric in the $(t,z)$ directions to be diagonal,
\be
g_{ab} dx^a dx^b = -g_{tt} dt^2 + g_{zz} dz^2.
\ee
We are now interested in accessing the physics of this equilibrium flow from a Euclidean partition function. Thus, consider the Wick rotation of \eqref{lor} to Euclidean signature. We identify Euclidean time with period $T_0^{-1}$ and we identify the $z$ coordinate with period $L_0$. We further expand all quantities in derivatives in the remaining $x^i$ directions. 

What is the gauge-invariant data on which the partition function can depend? To lowest order in derivatives, it can depend on the proper distances around the time and $z$ cycles:
\begin{align}
\frac{g_{tt}(x^i)}{T_0} , && g_{zz}(x^i)L_0,
\end{align}
as well as the gauge-invariant Wilson ``sheet'', corresponding to the integral of the background source $B_{\mu\nu}$ over the $(t,z)$ torus:
\be
\sW(x^i) \equiv \int B_{tz}(x^i) dt dz = B_{tz}(x^i)\frac{L_0}{T_0} .
\ee
Now to the lowest order in derivatives the partition function will take the form:
\begin{align}
&\log Z \equiv W \nn
=&\, \frac{1}{T_0}\int d^{3}x \sqrt{-g} {\mathcal{P}}\le(T_0\sqrt{g^{tt}}, L_0 \sqrt{g_{zz}}, B_{tz}\sqrt{g^{tt}g^{zz}}\ri),
\end{align}
where we have chosen to take as an independent variable the Wilson sheet normalized by the proper distances, and where everything depends on the transverse coordinates $x^i$. Now if we assume that the $z$ coordinate distance $L$ is much bigger than the other scales in the problem we can neglect the dependence of the partition function on it, and conclude that
\be
W(L \to \infty) = \frac{1}{T_0}\int d^{3}x \sqrt{-g} {\mathcal{P}}\le(T_0 \sqrt{g^{tt}}, B_{tz}\sqrt{g^{tt}g^{zz}}\ri).
\ee
For notational convenience we will denote the two arguments of $\sP$ by $\al$ and $\beta$ respectively. Now we use the definitions of the stress-energy tensor in terms of functional derivatives with respect to sources:
\begin{align}
T_{\mu\nu}(x) &= -\frac{2 T_0}{\sqrt{-g}} \frac{\delta W}{\delta g^{\mu\nu}(x)} ,\\
J^{\mu\nu}(x) &= \frac{T_0}{\sqrt{-g}}\frac{\delta W}{\delta b_{\mu\nu}(x)},
\end{align}
to conclude:
\begin{align}
T^{tt} & = -g^{tt}\le(\sP - \sqrt{g^{tt}}T_0 \p_{\alpha}\sP - \sqrt{g^{tt} g^{zz}} B_{tz}\p_{\beta} \sP\ri) ,\\
T^{ij} & = \sP g^{ij} ,\\
T^{zz} & = g^{zz}\le( \sP - \sqrt{g^{tt} g^{zz}}B_{tz} \p_{\beta} \sP\ri), \\
J^{tz} & = \sqrt{g^{tt} g^{zz}} \p_{\beta} \sP .
\end{align}
These expressions have precisely the form of the ideal hydrodynamic stress-energy tensor postulated in \eqref{TJapp}, with the identifications:
\begin{align}
p(x) &= \sP(x) ,\\ 
\ep &= -\sP + \sqrt{g^{tt}}T_0 \p_{\alpha}\sP + \sqrt{g^{tt} g^{zz}} \p_{\beta} \sP ,\\ 
\rho  &=  \p_{\beta}\sP ,\\ 
\mu &= \sqrt{g^{tt} g^{zz}}B_{tz}  . 
\end{align}
If we further identify
\begin{align}
T = \sqrt{g^{tt}}T_0, && s = \p_{\alpha}\sP,
\end{align}
we find that the expression for the energy is actually the thermodynamic identity
\be
\ep + p = Ts + \mu \rho.
\ee
Note that the relation $dp = s dT + \rho d\mu$ is now automatically satisfied. 

It was shown in \cite{Banerjee:2012iz,Jensen:2012jh} that, in general, such considerations can be extended to higher orders in the derivative expansion, and provide an alternative route to deriving the constraints on transport that arise from the local form of the second law. It would be interesting to understand whether this is the case for the higher-form current studied in this paper.

\section{Hydrodynamics of a $1$-form current at zero temperature.} \label{hydro0}
Here, we discuss the physics of a zero-temperature standard hydrodynamics with a 1-form conserved current, rather than the 2-form theory (cf. Section \ref{sec:T0}). As this is a simpler version of the theory discussed in Section \ref{sec:T0}, we will be somewhat brief. 

The theory has a conserved stress-energy tensor and 1-form $U(1)$ current. The degrees of freedom are a normalized fluid velocity and the chemical potential $\mu$. At the ideal level the stress-energy tensor and current are given by
\begin{align}
T^{\mu\nu}_{(0)} = (\ep + p)u^{\mu} u^{\nu} + p\, g^{\mu\nu} ,&& j^{\mu}_{(0)} = \rho \,u^{\mu} ,\label{zeroT1form}
\end{align}
where the relevant thermodynamic relations are just as in \eqref{thermoZeroT}
\begin{align}
\ep + p = \mu \rho , && dp = \rho d\mu.
\end{align}
We note that there are a total of four degrees of freedom (in 4 spacetime dimensions). However, there are five equations of motion:
\begin{align}
\nabla_{\mu} T^{\mu\nu}_{(0)} = 0, && \nabla_{\mu} j^{\mu}_{(0)} = 0 .\label{appeom}
\end{align}
Thus, this system naively appears overconstrained. This is an illusion, as in fact the combination of equations
\be
\le(\nabla_{\mu} T^{\mu\nu}\ri)u_{\nu} + \mu \nabla_{\mu} j^{\mu} = 0 \label{1formT0cons}
\ee
holds off-shell, i.e. for any stress-energy tensor and current that are parametrized in terms of fluid variables as in \eqref{zeroT1form}. Thus, this equation is redundant, and we have four dynamical equations for four degrees of freedom. 

It is interesting to note that in the usual finite temperature theory of ideal hydrodynamics, \eqref{1formT0cons} is instead (off-shell) equal to the divergence of the entropy current. Thus, it is tempting to think of the vanishing of the entropy at zero temperature as the key principle that allows a consistent zero-temperature hydrodynamics. 

We do not describe solutions to the set of equations \eqref{appeom} in great detail, except to state that they admit a linearly dispersing sound mode with dispersion relation
\begin{align}
\om^2 = v_s^2 k^2 , && v_s^2 = \frac{\rho}{\mu \chi} ,&& \chi = \frac{d\rho}{d\mu} \ . 
\end{align}

It would be interesting to develop this theory further, e.g. to go beyond ideal hydrodynamics, or to understand whether it can be related to fluctuations of holographic finite density systems at zero temperature such as those studied in \cite{Edalati:2010hk,Edalati:2010pn}. We leave these questions to future work. 

\section{Classification of tensor structures in MHD at zero temperature}\label{app:T0Tensors}

In order to write all possible higher order (in derivatives) modifications of the constitutive relations (\ref{idealTJzeroT}), one needs to construct all possible tensors of the desired type out of the building blocks provided by the effective degrees of freedom. In our case, after linearization (\ref{linear}), these are

\be
\delta u^{a i}, \, u_{a b},\, \omega^a, q^i,\, \delta \mu.
\ee

Notice that we have used the same notation as in the main section of this paper where $SO(1,1)$ indices are given by $a, b, \ldots$ and $SO(2)$ indices by $i, j, \ldots$. the fact that $\delta u^{a i}$ is in a bi-fundamental representation is a direct consequence of the constraint (\ref{uNorm2}).

An expansion in derivatives is an expansion in $\omega^a$ and $q^i$. To first order, it is straightforward to see that no two index object can be constructed with the above ingredients. This implies that at $T=0$ we can't correct our ideal equations to first order. The situation becomes more interesting at second order.

As in standard hydrodynamics, the classification of tensor structures for higher derivative corrections is greatly simplified by making use of ideal equations of motion. Because this is a perturbative construction performed order by order, one can always use lower order equations of motion in constraining higher order terms.

The ideal equations of motion can be efficiently written by projecting them onto $SO(1,1)$ and $SO(2)$ vector equations. First, consider the conservation of the antisymmetric current $J^{\mu \nu}$:
\bea
SO(1,1) &:&~~ \chi \, \omega_b \, u^{b a} \, \delta \mu = \rho \, q_i \,\delta u^{a i} , \\ 
SO(2) &:&~~ \rho \, \omega_a \delta u^{a i} =0 . 
\eea
The first of the above equations states that any $SO(1,1)$ derivative of $\delta \mu$ can be expressed in terms of the derivatives of $\delta u^{a i}$. The second equation states that we cannot contract $\delta u^{a i}$ with $\omega^a$.

The conservation of $T^{\mu \nu}$ can also be decomposed and expanded to linear order to yield

\bea
SO(1,1) &:&~~ \mu \chi \, \omega^a \, \delta \mu = \mu \, \rho \, q_i \,\delta u^{b i} u_{b}^{\phantom{b} a} , \\ 
SO(2) &:&~~ \rho \, q^i \delta \mu = \mu \, \rho u^{a b} \omega_a \delta u_b^{\phantom{b} i} .
\eea
The first equation above is identical to the $SO(1,1)$ equation arising form $J^{\mu \nu}$ conservation. This was pointed out around (\ref{trivialeq}) and is a crucial feature that allows for the consistency of the equations. The $SO(2)$ equation implies that $SO(2)$ derivative of $\delta \mu$ can be written in terms of derivatives of $\delta u^{a i}$. A straightforward consequence is that by using the above remarks, we can disregard $\delta \mu$ completely in building our higher derivative corrections, as this is not an independent quantity.

One can also consider further constraints that appear at the level of two derivatives by considering the ``integrability" conditions:
\bea
0 &=& \omega^{[a} \, \omega^{b]} \delta \mu = \frac{\rho}{\chi} \, q_i \,\delta u^{c i} u_{c}^{\phantom{c} [a} \omega^{b]} , \\
0 &=& q^{[i} q^{j]} \delta \mu = \mu \, u^{a b} \omega_a \delta u_b^{\phantom{b} [i} q^{j]}, \\
q^{i} \omega^a  \frac{\delta \mu}{\mu} & =&   \frac{\rho}{\mu \chi} \, q_j \,\delta u^{b j} u_{b}^{\phantom{b} a} q^i  = u^{b c} \omega_b \delta u_c^{\phantom{c} i} \omega^a .
\eea
These constraints end up reducing the number of independent antisymmetric and bi-fundamental tensors.

Armed with the above constraints, we can classify all tensors quadratic in derivatives, (or quadratic in powers of $\omega^a$, $q^i$). Notice that because we build only linear structures and $\delta \mu$ has been excluded from the building blocks as a consequence of the ideal equations of motion, the tensor $\delta u^{a i}$ must appear exactly once in all tensor structures. Lastly, we consider a theory respecting the charge assignments of table \ref{tbl:disc}. This implies that all corrections to the energy momentum tensor $T^{\mu \nu}$ must be even under $(u_{\mu \nu} \rightarrow - u_{\mu \nu} , \delta u^{a i} \rightarrow - \delta u^{a i})$ , while corrections to the current $J^{\mu \nu}$ must be odd. 

Using these rules, we construct all possible scalars and tensors:

\begin{itemize}
\item (even) scalars: $\delta u^{a i} q_i u_a^{\phantom{a} b} \omega_b$, 
\item (even) symmetric traceless $SO(1,1)$ tensors: $2 q_i \, \delta u^{c i} \, u_{c}^{\phantom{c} (a} \,  \omega^{b)}  - \Omega^{a b}  q_i \, \delta u^{c i} \, u_{c d} \,  \omega^d$, 
\item (even) symmetric traceless $SO(2)$ tensors: $2 q^{(i} \, \delta u^{a  j) } \,  u_{a b} \,  \omega^b -  \Pi^{i j}  q_k \, \delta u^{a k} \, u_{a b} \,  \omega^b$, 
\item (even) bi-fundamental tensors: $q_j \delta u^{b j} u_{b}^{\phantom{b} a} q^i$, $\delta u^{b i} \, u_{b}^{\phantom{b} a} \omega^c \omega_c$, \, $\delta u^{b i} \, u_{b}^{\phantom{b} a} q^j q_j$ ,
\item (odd) antisymmetric $SO(2)$ tensors: none.
\end{itemize}

These structures match exactly the expressions (\ref{s1}-\ref{s6}). A few comments are important regarding details in the construction of the above structures. First, one could have guessed that a new symmetric $SO(1,1)$ tensor can be built by acting on the one above on both indices with $u^a_{\phantom{a} b}$. The tracelessness condition implies this new tensor is equal to the one presented above. Also, one could have naively constructed a fourth bi-fundamental tensor, but this one can be expressed in terms of the ones above by the ``integrability" constraints. Lastly, parity under  $(u_{\mu \nu} \rightarrow - u_{\mu \nu} , \delta u^{a i} \rightarrow - \delta u^{a i})$ is responsible for the absence of $SO(2)$ antisymmetric tensors. Even if the parity condition were to be relaxed, the constraints would still prohibit this structure.

\end{appendix}
\bibliographystyle{utphys}
\bibliography{all}

\end{document}